\documentclass[review]{elsarticle}

\usepackage{graphicx}
\usepackage{dcolumn}
\usepackage{pifont}
\usepackage{bm}
\usepackage{multirow}
\usepackage{amsmath}
\usepackage{float}
\usepackage{txfonts}
\usepackage[version=3]{mhchem}

\def\mathbi#1{\textbf{\em #1}}
\def\mtxt#1{\mbox{\scriptsize #1}}
\def\mathbi#1{\textbf{\em #1}}
\def\scbi#1{\textbf{\em \scriptsize #1}}

\journal{JPhys Energy}

\begin{document}
\title{Advances in Modelling and Simulation of Halide Perovskites for Solar Cell Applications}

\author[kimuniv]{Chol-Jun Yu\corref{cor}}
\ead{cj.yu@ryongnamsan.edu.kp}

\cortext[cor]{Corresponding author}

\address[kimuniv]{Faculty of Materials Science, Kim Il Sung University, Taesong District, Pyongyang, Democratic People's Republic of Korea}

\begin{abstract}
Perovskite solar cells (PSCs) are attracting great attention as the most promising candidate for the next generation solar cells. This is due to their low cost and high power conversion efficiency in spite of their relatively short period of development. Key components of PSCs are a variety of halide perovskites with ABX$_3$ stoichiometry used as a photoabsorber, which brought the factual breakthrough in the field of photovoltaic (PV) technology with their outstanding optoelectronic properties. To commercialize PSCs in the near future, however, these materials need to be further improved for a better performance, represented by high efficiency and high stability. As in other materials development, atomistic modelling and simulation can play a significant role in finding new functional halide perovskites as well as revealing the underlying mechanisms of their material processes and properties. In this sense, computational works for the halide perovskites, mostly focusing on first-principles works, are reviewed with an eye looking for an answer how to improve the performance of PSCs. Specific modelling and simulation techniques to quantify material properties of the halide perovskites are also presented. Finally, the outlook for the challenges and future research directions in this field is provided.
\end{abstract}

\begin{keyword}
Halide perovskites \sep Solar cells \sep Electronic structure \sep Phonon \sep Defects \sep Interface
\end{keyword}

\maketitle

\section{Introduction}
As often highlighted on the public media, the issue of climate change and global warming is one of the most serious global challenges that human beings face nowadays~\cite{Paris}. This is an inevitable result from mass burning of fossil fuels by ourselves to run factories, drive cars and operate buildings during the past decades. In fact, this brought a release of a vast amount of greenhouse gases, such as carbon dioxide (CO$_2$) and methane (CH$_4$), and caused another challenge for energy due to the exhaustion of fossil fuel resources. To mitigate the catastrophic effects of global warming by reducing greenhouse gas emission and ensure the sustainable energy supply, more and more peoples and authorities express their growing interests in the renewable and clean (or low-carbon) energy sources, such that the majority of national and global energy policies include an exploitation of clean energy sources. Among several kinds of clean energy sources available on the globe, solar energy is the most promising source due to its abundance and eternal nature. Therefore, it is required to develop efficient techniques and devices for harnessing solar energy, which can contribute to resolving the environmental and energy challenges~\cite{Catlow10ptrsa}.

The most effective devices for harvesting solar energy should be undoubtedly solar cells invented in 1960s, which convert solar light directly into electricity by photovoltaic (PV) effect. So far, several kinds of solar cells have been developed, and can be found in the market; the market share is as 69.5\% for single crystalline silicon ($sc$-Si), 23.9\% for polycrystalline silicon ($pc$-Si) and 6.6\% for thin-film polycrystalline CdTe solar cells in 2015~\cite{Chu17nm}. In fact, the most important factors determining the competence of solar cell in the market are power conversion efficiency (PCE), fabrication cost and device stability. According to the report of Green et al.~\cite{Green17ppra}, certified PCEs of conventional PV modules are 28.8\% for $sc$-GaAs, 24.4\% for $sc$-Si, 19.9\% for $pc$-Si, 18.6\% for CdTe or Cu-In-Ga-Se (CIGS) alloy, and 13\% for quantum dot and dye-sensitized solar cells. For these conventional solar cells, it is difficult to get a high efficiency, a low cost and a high stability simultaneously; when the efficiency increases, the fabrication cost also increases. That is why $sc$-GaAs solar cells with the highest cost and high stability and dye-sensitized solar cells with the lowest efficiency and low stability could not be commercialized. For the most widely used Si solar cells, however, both the efficiency and cost are not satisfactory enough to become competitive with fossil fuels in the electricity market. Therefore, it is our ardent desire to invent an innovative new type of solar cell with high efficiency, low cost and high stability together.

In accordance to this desire, perovskite solar cells (PSCs) have emerged with an initial PCE of 3.8\% in 2009~\cite{Kojima09jacs}, and evolved rapidly with their PCEs to over 10\% within a few years~\cite{Im11ns,Lee12s,Kim12sr,Burschka13n}. After several years of intensive and extensive research, the efficiency arrives at over 22\%~\cite{Zhou14s,Jeon15n,Yang17s}, which is an astonishing breakthrough in the PV history when compared with the development periods of the conventional solar cells mentioned above. Due to simple fabrication processes (low processing cost) and highly abundant raw materials (low materials cost)~\cite{Snaith13jpcl,Park13jpcl}, moreover, their fabrication cost is a third time as low as  $c$-Si solar cells, resulting in a short energy payback time and low overall CO$_2$ emission. Such huge success is mostly originated in their key components, halide perovskites with a chemical formula of ABX$_3$, where A and B are the monovalent (organic or inorganic) and divalent (metallic) cations and X is the halide anion, respectively. These halide perovskites have been used as a solar light absorber and/or a charge carrier transmitter due to their optimal properties for these aims~\cite{Li18jmca,Chen18jpcl,Huang17nrm,Mohammed17rser,Mahmood17ra,Hoye17cm,Xiao16mser,Park16ne,Zuo16as,Sum14ees}. For instance, the archetype methylammonium lead iodide (CH$_3$NH$_3$PbI$_3$, MAPI hereafter), which was firstly hired in PSCs, has appropriate band gaps, optimal photoabsorption coefficients, weak exciton binding energies, high charge carrier mobilities, and long carrier diffusion lengths. However, there are still several fundamental issues to be addressed for commercialization of PSCs: (i) high stability and long lifetime~\cite{Wang16semsc,Berhe16ees,Li16ra,Niu15jmca,Manser16acr}, (ii) scalability for large area modules~\cite{Li18nrm} and (iii) low toxicity~\cite{Chatterjee18jmca}.

Atomistic modelling and simulation of materials become essential in materials science, as they can provide valuable insights for known material processes and properties, and further predict the shortest route for finding new functional materials that meet the requirements~\cite{Catlow10ptrsa,Butler16cmd,Luo15jmca}. In particular, first-principles methods based on quantum mechanics can quantitatively determine most of material characteristics of crystalline solid, such as lattice constants, electronic structure, linear response properties, and transport properties, with a typical accuracy of 1\% relative error to the experiment by using input data of only atomic informations and known (or hypothetical) lattice structure. Being an interdisciplinary subject, the power of computational materials science can be supported by two aspects: (i) in hardware, rapid and ever increase of processor speed and memory capacity, and (ii) in software, continuous progress of simulation algorithms and materials theory. As such, the state-of-the-art atomistic modelling and simulation methods such as density functional theory (DFT), many-body perturbation theory (MBPT), and pair-wise interatomic potential molecular dynamics (MD) have been successfully applied to the halide perovskites. Recently, there exist several reviews on theoretical works of halide perovskites: short reviews on the nature of chemical bonding~\cite{Walshjpcc15} and on the electronic and ionic motions~\cite{Frostacc16}, perspectives on the entropy~\cite{Butler16cs} and on hybrid halide perovskites~\cite{Whalleyrev17}, and reviews focusing on MD simulations~\cite{Mattoni16jpcm} and on optoelectronic properties~\cite{Yin15rev,Manser17rev}.

In this review, we inclusively discuss recent progress in first-principles atomistic modelling and simulation of halide perovskites in both hybrid organic-inorganic and purely all-inorganic forms. We focus our special attention on how the calculated material properties can be linked up with the performance of PSCs. Specific modelling and simulation techniques to elucidate material properties of halide perovskites are also discussed. We consider the following material properties: (i) crystalline structures, (ii) electronic structures and optical properties, (iii) phonon dispersions and material stability, (iv) defect physics and ionic diffusion, and (v) surface and interface. The accuracies of calculations adopted in each work is also on discussion in comparison with the available experimental data. Finally we present the challenges to modelling and simulation of halide perovskites and future research directions.

\section{Crystalline structures and polymorphism}
The first step of solid-state modelling and simulation is to define the Bravais lattice (lattice constants and angles) and structural factor (atomic coordinates) of unit cell. This can be done simply from experimental data such as X-ray diffraction (XRD) and scanning or transmission electron microscopy (SEM or TEM) with elemental analysis (e.g. energy dispersive X-ray spectroscopy: EDX or EDS) for a known material, where data mining is a crucial tool to analyze large data sets~\cite{Hautier13nc} and to predict the structure-property cartograms~\cite{Isayev15dm}. Then, the initial structure is locally refined by performing structural optimization, which can provide a preliminary accuracy estimation of adopted theoretical method by comparing the determined lattice constants with experimental data, and a supplementary information of elasticity (bulk modulus) by postprocessing the calculated data of total energy vs unit cell volume. If the structure is unknown, the crystal structure prediction should be performed in different ways: (i) simply making a set of hypothetical crystalline structures and then comparing their total energies, and (ii) global structure optimization using a variety of genetic and evolutionary algorithms~\cite{Oganov11acc,Pickard11go}. Once the refined structure is obtained, we can perform the analysis of chemical bonding by measuring the bond length and bond angles, and proceed further calculations for material properties, getting the relation of structure-property.

\subsection{Geometric factors for 3D halide perovskites}
In ABX$_3$ halide perovskite structure, the large monovalent cation A is bonded with the nearest twelve X halide anions, forming an AX$_{12}$ cuboctahedron, while the smaller divalent metal cation B forms corner sharing BX$_6$ octahedra in a three-dimensional (3D) framework. It should be noted that A cation can be organic, forming hybrid organic-inorganic halide perovskites, and inorganic, leading to all-inorganic halide perovskites. At a glance, it seems that there are a great number of combinatorial cases simply with monovalent A and divalent B cations based on the periodic table. However, there are some geometric factors restricting the formation of stable 3D halide perovskite structure. The first thing is the Goldschmidt tolerance factor~\cite{Goldschmidt1926}, $t$, defined as,
\begin{equation}
t=\frac{r_A+r_X}{\sqrt{2}(r_B+r_X)} \label{eq_gold}
\end{equation}
where $r_A$, $r_B$ and $r_X$ are the ionic radii of A, B, and X ions, respectively. It was empirically known that with $t=1$ for a perfect cubic lattice, $0.8\leq t \leq 1.0$ for most stable perovskite structures with tetragonal, orthorhombic and rhombohedral (or trigonal) lattices. Otherwise, nonperovskite structures are formed; hexagonal structure is formed when $t>1$ and different structures are found when $t<0.8$~\cite{Stoumpos15acr,Travis16cs}.

\begin{figure}[!th]
\begin{center}
\includegraphics[clip=true,scale=0.5]{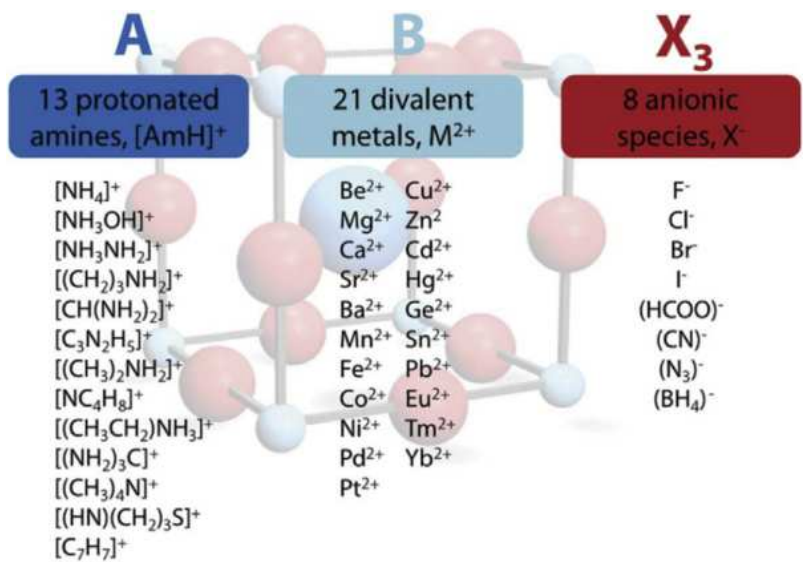}
\caption{\label{fig_gtf}Based on the Goldschmidt tolerance factor calculation for 2352 amine-metal-anion compounds, 562 organic anion-based and 180 halide-based perovskites are selected to have $0.8\leq t \leq 1.0$. Reproduced from~\cite{Kieslich15cs}. CC BY 3.0.}
\end{center}
\end{figure}

Providing the halide ions (X = F$^-$, Cl$^-$, Br$^-$, I$^-$), typical examples of B-site metal cations are B = Pb$^{2+}$, Sn$^{2+}$, Ba$^{2+}$, Ge$^{2+}$, and monovalent A-site cations include organic CH$_3$NH$_3^+$ (MA), HC(NH$_2$)$_2^+$ (FA) and CH$_3$CH$_2$NH$_3^+$ (EA) cations and inorganic Cs$^+$ and Rb$^+$ cations. By using the calculated tolerance factors, Kieslich et al.~\cite{Kieslich15cs} demonstrated that 2352 amine-metal-anion compounds are possible, of which 180 halide compounds (and 562 organic anion-based) can be stable perovskites with $0.8<t<1.0$ (Fig.~\ref{fig_gtf}). These are too many cases to consider in computational work. In such situation, another constraint factor should be necessary to define.

\begin{figure}[!th]
\begin{center}
\includegraphics[clip=true,scale=0.25]{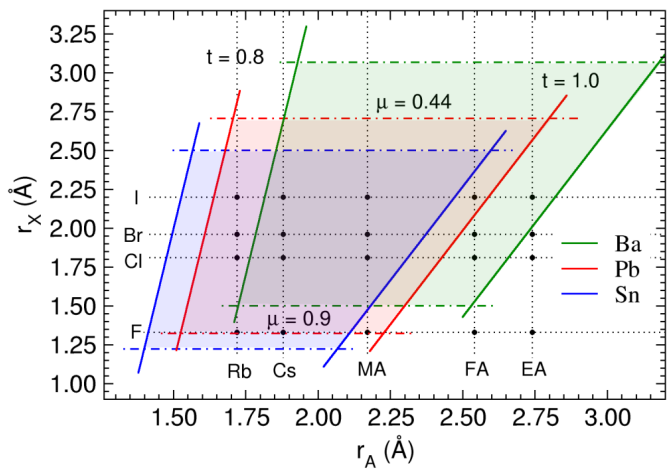}
\caption{\label{fig_oct}2D map of parameter space for stable perovskites based on B-site metals of Pb, Sn and Ba. Ionic radii of A cations and X anions are marked along the abscissa and ordinate, respectively. Solid and dashed lines indicate limits of tolerance and octahedral factors. Adapted from~\cite{Manser17rev} with permission of the American Chemical Society. }
\end{center}
\end{figure}

The second constraint is known as the octahedral factor, defined as an ionic radii ratio between B cation and halide anion, $\mu=r_B/r_X$~\cite{CLioctfac08}. This reflects the stability of BX$_6$ octahedron, and should be in the range of $0.44\leq\mu\leq0.9$ for a stable perovskite. Together with the tolerance factor, the octahedral factor can provide a constraint to select stable halide perovskites among many possible cases. For instance, we can draw a 2D map of ionic radii of A cations (Rb$^+$, Cs$^+$, MA$^+$, FA$^+$, EA$^+$) and X anions (F$^-$, Cl$^-$, Br$^-$, I$^-$), which are marked along the abscissa and ordinate in Fig.~\ref{fig_oct}, for typical B-site metal cations of Pb$^{2+}$, Sn$^{2+}$ and Ba$^{2+}$. Their ionic radii with proper coordination numbers (CN = 12, 6, 6 for A, B, X ions, respectively) are provided in Refs.~\cite{Shannon76ion,crchand2004}. In Fig.~\ref{fig_oct}, we can see the intersections located within the polygonal (shaded) region formed by tolerance limits of these constraint factors, which indicate a potentiality of stable perovskite formation with corresponding A, B, X elements, and those outside the region, which indicate a fail of perovskite formation. It should be noted that there is an ambiguity to define ionic radii due to the nonsphericity of organic molecular A cation and reduced electronegativity of heavy halide anions, and thus some modifications have been proposed, such as effective molecular A cation radii~\cite{Kieslich15cs} and modified B-site metal cation radii from empirical bond lengths~\cite{Travis16cs}.

\subsection{Octahedral tilting or distortion and polymorphism}
Once a certain compound with ABX$_3$ formula is proven to form a stable halide perovskite due to its proper tolerance and octahedral factors, we should also consider a possibility of modification in the shape of BX$_6$ octahedra. In fact, BX$_6$ octahedra can be tilted or distorted because of slight rotations or displacements of constituent atoms or molecules at finite temperature (and possibly pressure). It is worth noting that the octahedral tilting is known to be associated with antiferroelectricity, while the octahedral distortion caused by cation off-centering yields a ferroelectricity (i.e. creation of spontaneous electric polarization) that can enhance charge carrier separation and allow the photovoltage to exceed the band gap by thousands of times~\cite{Beecher16ael,Leguy15nc}. Such octahedral tilting or distortion may lose a symmetry of crystalline unit cell, being closely related with phase transition driven by temperature~\cite{Glazer1972}. In general, the archetype cubic perovskite structure is observed at high temperature, and when decreasing temperature, lower symmetry phases are found in the order of tetragonal$\rightarrow$orthorhombic$\rightarrow$monoclinic and/or rhombohedral.

\begin{figure}[!th]
\begin{center}
\includegraphics[clip=true,scale=0.45]{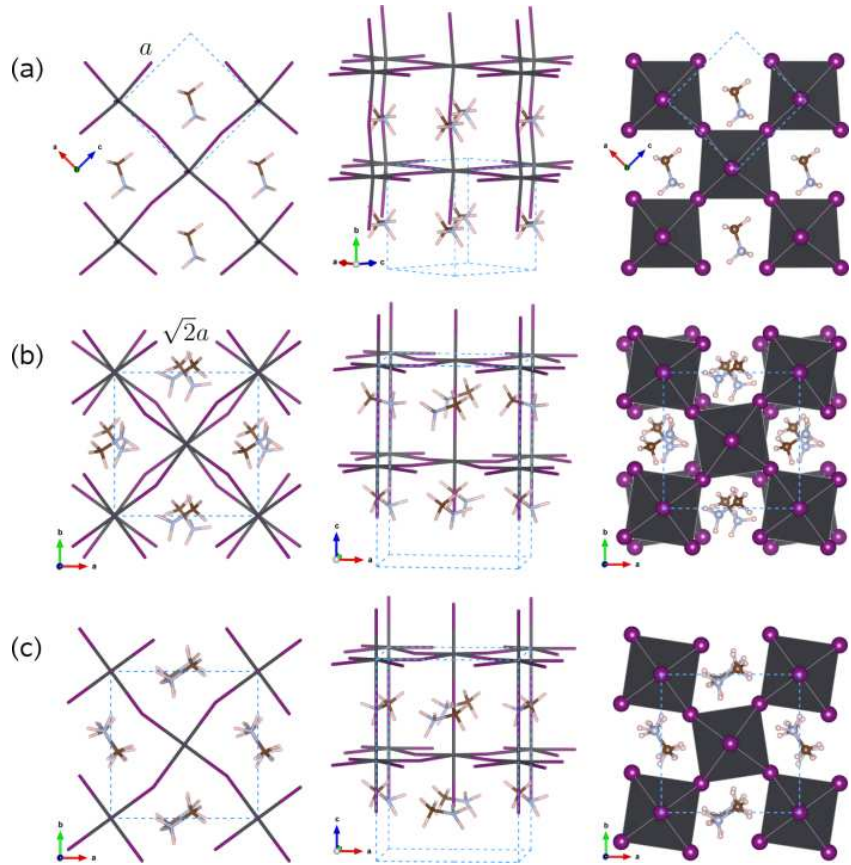}
\caption{\label{fig_mapimod}Atomistic modelling of MAPI with (a) cubic, (b) tetragonal, and (c) orthorhombic phases. Left and middle panels are for stick top and side views, and right panel is for polyhedral top view. Brown, pale pink, pale blue, black and violet colours represent C, H, N, Pb, and I atoms, respectively. Dashed blue lines indicate the corresponding unit cells.}
\end{center}
\end{figure}

For the typical case of hybrid organic-inorganic halide perovskite MAPI, which has been playing the major role in advancing PSC technology, its cubic $\alpha$-phase with space group $Pm$\={3}$m$ is only found at high temperature, and phase transitions to tetragonal $\beta$-phase ($I4/mcm$) occurs at 327.4 K, and to orthorhombic $\gamma$-phase ($Pna2_1$) at 162.2 K~\cite{Poglitsch87jcp}. In this case of MAPI, the orientation of MA cation is of great importance in determining the octahedral tilting and the crystalline phase~\cite{Leguy15nc,Brivio13apl}. The organic MA$^+$ cation has two modes of reorientation: (i) methyl and/or ammonium rotation around the C$-$N axis and (ii) whole molecular rotation of the C$-$N axis itself~\cite{Leguy15nc}. In the cubic phase, three different orientations are identified as $<$100$>$, $<$110$>$, and $<$111$>$ for the C$-$N axis, which lower lattice symmetry from the cubic $Pm$\={3}$m$ to the pseudo-cubic $Pm$ and $R3m$ phases, and the $<$110$>$ model was found to be the most energetically stable configuration by {\it ab initio} MD~\cite{Brivio13apl} and structural optimization~\cite{yucj16jms} in agreement with the experiment~\cite{Leguy15nc}. By performing {\it ab initio} MD simulations, the activation barriers were calculated to be 117 meV for reorientations of the axis~\cite{Mosconi14pccp} and 13.5 meV for ion rotation~\cite{Frost14aplm,Frost14nl}. For the tetragonal phase, the C$-$N bonds in MA distribute in a vertical way, while for the orthorhombic phase in a parallel way, as shown in Fig.~\ref{fig_mapimod}. In these non-cubic phases, their unit cells ($\sqrt{2}a\times\sqrt{2}a\times2a$ supercell) contain 4 formula units (48 atoms), being larger than the cubic phase with one formula unit (12 atoms)~\cite{WGeng14jpcc}.

Temperature-driven phase transitions were also observed for all-inorganic halide perovskite ABX$_3$ (A = Cs, Rb; B = Pb, Sn; X = F, Cl, Br, I) (see Refs.~\cite{Yang17jpcl} and therein). For CsSnI$_3$ typically, phase transitions from the perovskite cubic $\alpha$-phase ($Pm$\={322}$m$) upon decreasing temperature occur to the tetragonal $\beta$-phase ($P4/mbm$) at 446 K and to orthorhombic $\gamma$- and non-perovskite Y-phases ($Pnma$) at 373 K~\cite{Yamada91cl}, while for CsPbI$_3$ phase transition occurs only to the non-perovskite orthorhombic Y phase at 583 K~\cite{Trots08jpcs}. It is worth noting that for CsSnI$_3$ two orthorhombic structures coexist at room termperature, of which the yellow Y phase characterized by an edge-connected 1D double-chain is not useful for solar energy application~\cite{Silva15prb}. Unlike the hybrid perovskite, the all-inorganic halide perovskites do not have rotational molecular disorder in A-site, and instead the flexibility associated with the inorganic octahedra network solely causes the phase transition. By performing the phonon calculations for CsBX$_3$ (B = Pb, Sn) in cubic phase, Yang et al.~\cite{Yang17jpcl} revealed that these inorganic halide perovskites exhibit lattice instabilities in the cubic phase, confirmed by phonon soft-mode (imaginary frequencies) at the Brillouin zone boundary, and thus the octahedral tilting and distortion that is by definition antiferroelectric in nature can occur spontaneously. A quantitative picture for phase diversity can be obtained by calculating free energy through phonon calculation, which will be discussed later.

\subsection{Low dimensional and double lead-free halide perovskites}
With respect to the environmental impact, there is a concern on the toxicity of lead in the major halide perovskites such as MAPI and CsPbI$_3$, promoting an extensive search for alternatives. Although there has been extensive research to try simple substitution of Pb$^{2+}$ in ABX$_3$ structure with alternative divalent cations, it was turned out to be challenge. There is also research interest in another approach to search for multivalent elements. This resulted in low dimensional perovskite structures such as A$^{\mtxt{I}}_3$B$^{\mtxt{III}}_2$X$_9$ (``3-2-9'') layered (2D), A$^{\mtxt{I}}_3$B$^{\mtxt{II}}$X$_5$ (``3-1-5'') single chained (1D) and A$^{\mtxt{I}}_4$B$^{\mtxt{II}}$X$_6$ (``4-1-6'') isolated octahedra (0D), and double perovskite structures A$^{\mtxt{I}}_2$B$^{\mtxt{I}}$B$^{\mtxt{III}}$X$_6$ (``2-1-1-6'')~\cite{Yan18jmca,Maughan18cm}.
Xiao et al.~\cite{Xiao17mh} has reported their recent work for Pb-free halide perovskites with such different structural dimensionalities, which have a wide range of band gaps and PCEs of those-based solar cells. In particular, 3D double perovskites such as Cs$_2$AgBiX$_6$ (X = Br, Cl), called elpasolites, have received much attention as promising absorber materials~\cite{Savory16ael,McClure16cm,FWei16mh,ZDeng16jmca,Filip16jpcl}. Volonakis et al.~\cite{Volonakis16jpcl} has reported a series of double perovskites with B$^{\mtxt{I}}$ = Cu, Ag, Au and B$^{\mtxt{III}}$ = Bi, Sb, predicting their band gaps in the range from 0.5 to 2.7 eV for Bi system and from 0.0 to 2.6 eV for Sb system. So far, there is no report for double perovskite-based solar cells, which might be due to difficulties in synthesizing uniform thin films of the correct phase and chemical composition.

\subsection{Solid solutions}
There is an ever-increasing need towards higher performance of solar cells, represented by efficiency and stability. Although the PCE of MAPI-based solar cells reaches as high as 22\%, being comparable with $sc$-Si solar cells, their instability in air is relatively low, which could be major barrier to commercialization. Making solid solutions or mixing at A-site and/or X-site with two or more similar elements is an efficient way to satisfy the need of enhancing the performance by increasing the stability, improving the charge carrier transport, and tuning the band gap~\cite{Ono17aami,Chatterjee18jmca}. Experimentally, it is much easier to realize such mixing than double perovskites. There have been several first-principles investigations for perovskite solid solutions such as hybrid perovskites, MAPb(I$_{1-x}$Br$_x$)$_3$~\cite{yucj16jms,Brivio16jpcl,Mosconi13jpcc,yucj16prb}, MAPb(I$_{1-x}$Cl$_x$)$_3$~\cite{yucj17jps} and (Cs/MA/FA)PbI$_3$~\cite{Saidaminov18ne}, and inorganic perovskites, CsPb(X$_{1-x}$Y$_x$)$_3$ (X, Y = I, Br, Cl)~\cite{Yin14jpcl} and (Rb$_x$Cs$_{1-x}$)SnI$_3$~\cite{Jung17cm}. When mixing larger I$^-$ anion with smaller Br$^-$ or Cl$^-$ anion, the crystalline lattice can be reduced and a phase transition from the tetragonal to the cubic can occur at room temperature. Meanwhile, mixing MA$^+$ cation with larger FA$^+$ cation can tune the octahedral tilting with the proper Goldschmidt tolerance factor.

\begin{figure}[!th]
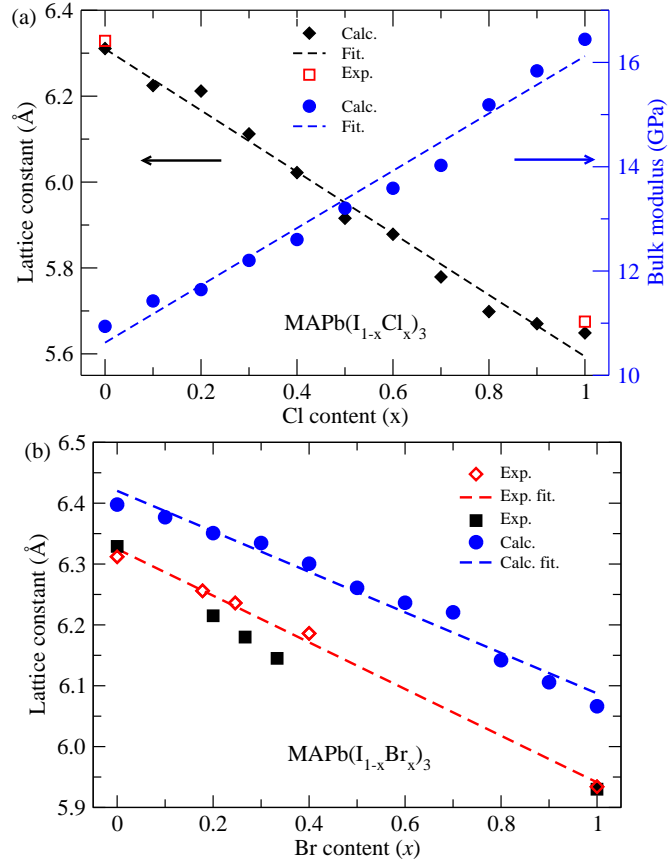

\begin{center}
~~~~~\includegraphics[clip=true,scale=0.5]{fig4a.eps} \\
\includegraphics[clip=true,scale=0.5]{fig4b.eps}
\caption{\label{fig_vcalat}Cubic lattice constants as a function of mixing content in (a) MAPb(I$_{1-x}$Cl$_x$)$_3$, reprinted figure with permission from~\cite{yucj17jps}, Copyright 2017 by the Elsevier B.V., and in (b) MAPb(I$_{1-x}$Br$_x$)$_3$, calculated by VCA approach, reprinted figure with permission from~\cite{yucj16prb}, Copyright 2016 by the American Physical Society.}
\end{center}
\end{figure}
To make modelling of solid solutions, we can use supercell (SC) method at the sacrifice of high computational cost. When employing the pseudopotential plane wave (PP-PW) method, the virtual crystal approximation (VCA) approach can be used alternatively to reduce the computational cost, but the accuracy or reliability should be checked carefully before going into the main calculation. In the VCA approach, the virtual atom is introduced and its pseudopotential is constructed by averaging the relevant potentials and wave functions of constituent atoms~\cite{yucj07jpcm,Iniguez03prb}. There are some technical problems in these methods; random distribution of alloying atoms for SC method, while local effect from the interaction between the constituent atoms for VCA method. It has been proven that the VCA approach can produce the reliable tendency of lattice constants in cubic phase as a linear function of mixing content, i.e. Vegard's law, for hybrid halide perovskites~\cite{yucj16prb,yucj17jps} (Fig.~\ref{fig_vcalat}).

\section{Electronic and optical properties}
Since the halide perovskites are typically used as a light absorber in most applications, the electronic and optical properties are of primary importance. After modelling the crystalline solid using a unit cell or supercell and refining their cell by performing the structural optimization, the electronic energy band structure and the corresponding density of states (DOS) can be obtained by non-SCF calculation. Postprocessing the energy band produces directly the effective masses of charge carriers (conductive electrons and holes). Other optical properties such as exciton binding energy and photoabsorption coefficients are determined by using the frequency-dependent dielectric constants, which can be routinely calculated within the density functional perturbation theory (DFPT)~\cite{Baroni01rmp,Sharma03prb,Sharma04ps} without or with the effect of electron-hole interaction by using the Bethe-Salpeter approach~\cite{Onida02rmp}.

\subsection{Spin-orbit coupling and many-body effects}
\begin{figure}[!th]
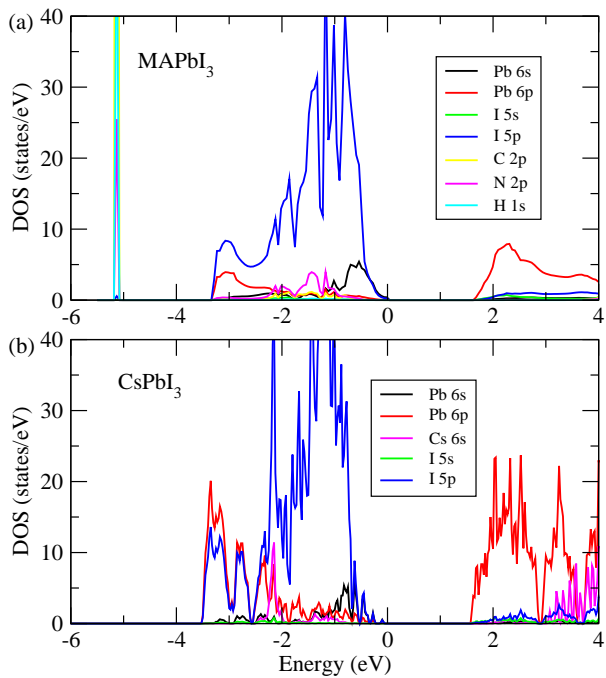

\begin{center}
\includegraphics[clip=true,scale=0.5]{fig5a.eps}
\includegraphics[clip=true,scale=0.5]{fig5b.eps}
\caption{\label{fig_dos}Partial density of states (DOS) for (a) MAPbI$_3$ and (b) CsPbI$_3$ in cubic phase, calculated by pseudopotential plane wave method. The valence band maximum is set to be zero.}
\end{center}
\end{figure}

In general, standard DFT exchange-correlation (XC) functionals within local or semi-local approximations such as local density approximation (LDA) and generalized gradient approximation (GGA) severely underestimate the band gap of inorganic semiconductors, due to the inherent nature of DFT as ground state theory and the artificial self-interaction between electrons, although they can provide reliable structures and stabilities. Surprisingly for MAPI, the GGA functionals (PBE~\cite{PBE96prl} or PBEsol~\cite{PBEsol08prl}) can yield a band gap in excellent agreement with experimental values within $\pm$0.1 eV~\cite{Mosconi13jpcc,yucj16prb,yucj17jps,Umebayashi03prb}. This is due to a fortuitous error cancellation between the GGA underestimation and the overestimation by the lack of spin-orbit coupling (SOC)~\cite{Even14jpcc,Even13jpcl,Motta15nc,Giorgi13jpcl,Azarhoosh16aplm}, which inversely indicates an importance of relativistic effects in halide perovskites. From the analysis of partial density of states (PDOS) for halide perovskites ABX$_3$, the valence band maximum (VBM) consists of an antibonding coupling of X $p$-states and B $s$-states, while the conduction band minimum (CBM) is dominated by B $p$-states (see Fig.~\ref{fig_dos}). It should be noted that for MAPI the major of occupied molecular orbitals of MA is found deep ($\sim$5 eV) below the VBM and the minor (but not negligible) contribution is found $\sim$0.5 eV below the VBM, indicating the hydrogen bonding interaction between the organic MA moiety and the inorganic PbI$_6$ octahedra~\cite{Umari14sr}. In APbX$_3$ perovskites the heavy Pb atom has a significant SOC effect of lowering the CBM, whereas the lighter Sn atom has a relatively weak SOC effect, resulting in again large underestimation of band gap by GGA functionals for ASnX$_3$ perovskites~\cite{Chung12jacs,Borriello08prb}. Moreover, the GGA functionals cannot accurately describe dispersion of the valence band even in Pb-based perovskites.

\begin{figure}[!th]
\begin{center}
\includegraphics[clip=true,scale=0.18]{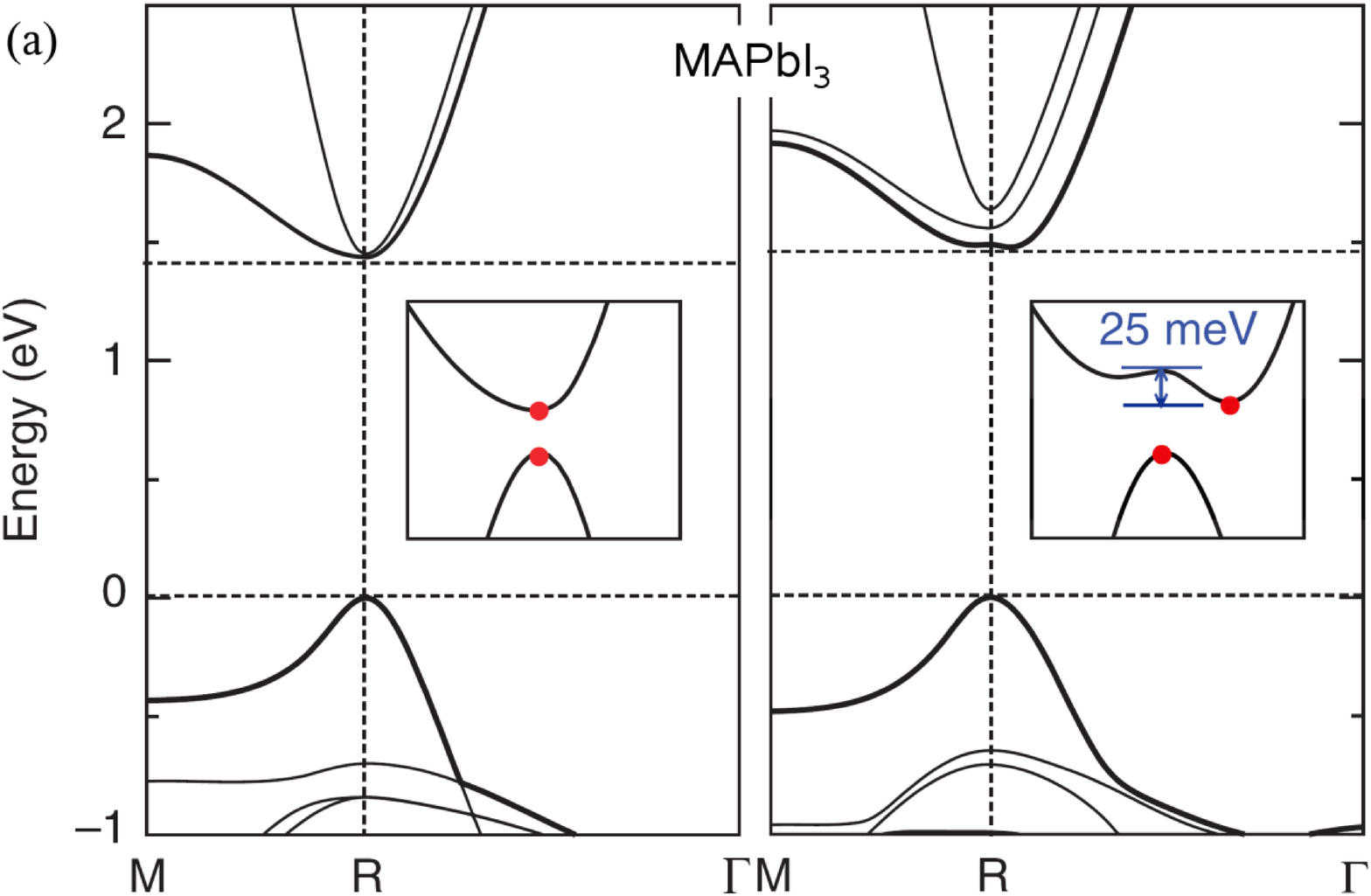}
\includegraphics[clip=true,scale=0.1]{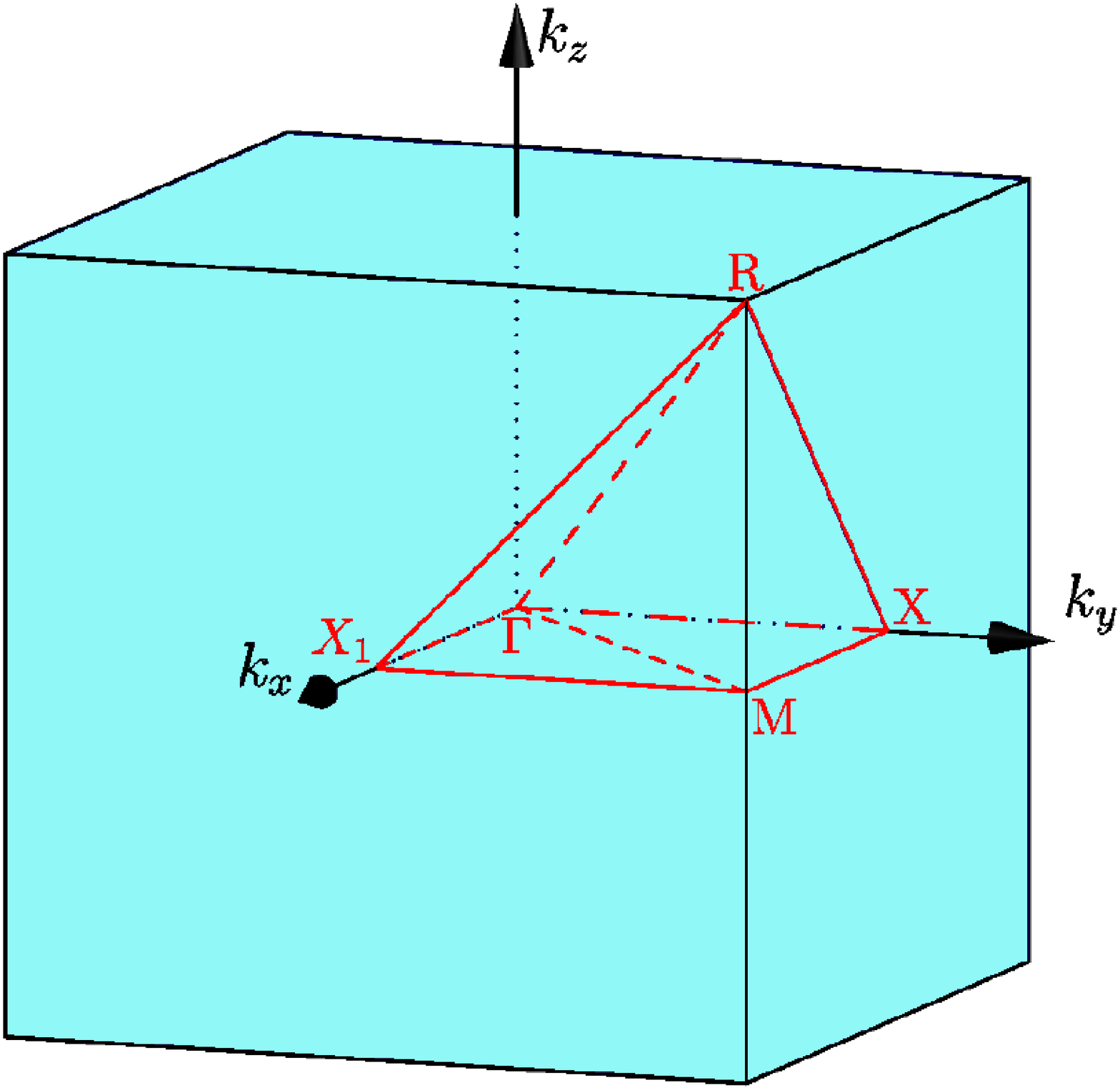}
\includegraphics[clip=true,scale=0.525]{fig6b.eps}
\caption{\label{fig_band}(a) Electronic energy band of hybrid iodide perovskite MAPbI$_3$ in cubic phase around R point without (left panel) and with SOC (right panel). Adapted from~\cite{Motta15nc}. CC BY 4.0. (b) Band structure of all-inorganic iodide perovskite CsPbI$_3$ in cubic phase without (dashed lines) and with SOC (solid lines). VBM is set to be zero. High symmetry points and lines in the Brillouin zone are shown.}
\end{center}
\end{figure}

The relativistic effect can be approxmated by first-order scalar relativistic (SR) and higher order SOC contributions. Umari et al.~\cite{Umari14sr} demonstrated that SR-DFT barely changes the band obtained by DFT, while SOC-DFT largely lowers the conduction bands for hybrid iodide perovskites, MAPbI$_3$ and MASnI$_3$. When the SOC effect is in consideration, splittings of the conduction bands, known as Rashba effect, occurs as a result of interaction between the magnetic moment (spin) of the electron and the local electric field~\cite{Kepenekian15an}. In fact, this electromagnetic force displaces electrons in the momentum space, acting on up and down spins in opposite direction. For the case of cubic halide perovskites, neglecting SOC yields direct band gaps at the band edge points of R and M in the Brillouin zone, as shown in Fig.~\ref{fig_band}. When SOC is turned on, both the valence and conduction bands around R split into symmetrical valleys for the case of MAPI (see Fig.~\ref{fig_band}(a)). More importantly, since the splitting is much more evident in the Pb 6$p$ conduction band compared with the I 5$p$ valence band, the CBM slightly shifts along the R$\rightarrow\Gamma$ line, resulting in the vertical energy difference of 25 meV between the CBM and VBM at R point~\cite{Motta15nc}. Thus the band gap changes into the indirect mode, which affect the function of light absorber as we discuss later. Motta et al.~\cite{Motta15nc} revealed that the orientation of MA molecular cation has the same effect on the band gap transition. For the all-inorganic halide perovskites like CsPbI$_3$, however, such relativistic spin-splitting does not occur due to a centre of inversion symmetry. Although the SOC effect causes the separation of Pb 6$p$ into $p_{1/2}$ and $p_{3/2}$, no splitting of the band extrema is observed at the high symmetry points, as shown in Fig.~\ref{fig_band}(b).

Since both the standard DFT and SOC treatment based on local or semi-local XC functionals are insufficient to describe the electronic structures of halide perovskites, it is necessary to adopt more delicate approaches. One way is to use a screened hybrid functional such as HSE06~\cite{HSE03jcp,HS04jcp}, which has a tendency of overestimating the band gap of hybrid and all-inorganic perovskites and thus by combining SOC treatment can give a reasonable estimation. More profoundly, as in other inorganic semiconductors, an accurate description of electronic structure can be provided by considering many-body interactions within $GW$ approximation~\cite{Umari14sr,Brivio14prb,Huang13prb} or random phase approximation (RPA)~\cite{Bokdam17prl}. Quasiparticle self-consistent $GW$ (QS$GW$) approach could yield the large overestimation of band gaps as 2.68$\sim$2.73 eV (1.55 eV) for MAPI (MASnI$_3$), while SOC+$GW$ by combining with SOC can give the closest band gaps of 1.67 eV (1.10 eV) to the experimental value of 1.60 eV (1.20 eV)~\cite{Umari14sr,Brivio14prb}. In addition, inclusion of SOC in $GW$ calculation increases band dispersion so significantly that the standard parabolic approximation can no longer be applied to the band extrema, which differs from experiment indicating parabolic band extrema in MAPI.

\subsection{Tuning band gap by substitution and mixing}
As the key factor of light absorber, the band gap of the halide perovskites can be tuned by different ways: (i) static volume change, (ii) temperature change, and (iii) chemical substitution. When decreasing the volume by acting a kind of small perturbation or by increasing pressure, the band gap was found to monotonically decrease in DFT calculation with an additional benefit of transition from indirect to direct band gap for MAPI~\cite{Wang17ees,Meloni16jmca}. Also, it was found that the out-of-phase band-edge states are stabilized as the lattice expands~\cite{Frost14nl}. Regarding the temperature effect, there is no theoretical work yet, although the band gap was observed in temperature-dependent photoluminescence to decrease with decreasing temperature from 1.61 eV at 300 K to 1.55 eV at 150 K for MAPI~\cite{Yamada14ape}. We focus on the third way of chemical substitution at each site of ABX$_3$, which is more genuine for materials design than the physical volume effect~\cite{Walshjpcc15,Amat14nl,Mao18jpcc}.

The A-site cation of MAPI can affect only indirectly on the band gap by changing the crystal structure due to its molecular orbitals far away from the VBM. When replacing MA cation with smaller molecular cation such as NH$_4^+$ and H$^+$, the band gap was found to lower by 0.3 eV with contraction of cubic lattice from 6.29 to 6.21 \AA~in QS$GW$+SOC calculation for NH$_4$PbI$_3$~\cite{Brivio14prb}, and to be less than 0.3 eV with a lattice constant of 6.05 \AA~for HPbI$_3$~\cite{Frost14nl}. It is important to recognize the relationship between the ionic radii and the local structure. In fact, both NH$_4$PbI$_3$ and HPbI$_3$ adopt nonperovskite chain or layer structures due to their lower tolerance factors than 0.81, which give rise to an instability of the octahedral networks with respect to tilting. Replacing MA with larger FA can also result in structural distortions~\cite{Pang14cm}. Meanwhile, substituting inorganic cation Cs$^+$ with a smaller ionic radius leads to a widening of band gap as 1.73 eV in cubic CsPbI$_3$~\cite{Eperon15jmca,Hendon15jmca,Brgoch14jpcc}, being suitable for top cell material in tandem cell. However, it is difficult to form its cubic phase at room temperature. In this situation, it is desirable to adopt the mixing technique, which can promise a synergy effect of enhancing stability against moisture and keeping efficiency. Mixing can be done in various ways such as MA/FA, Cs/MA, Rb/Cs~\cite{Jung17cm}, Cs/MA/FA~\cite{Saidaminov18ne}, and Rb/Cs/MA/FA~\cite{Duong17aem}. 

\begin{figure}[!th]
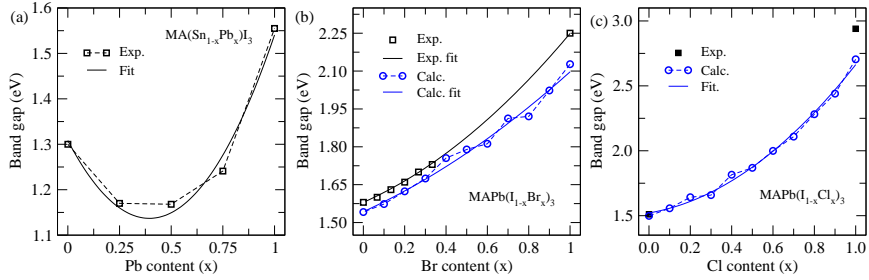

\begin{center}
\includegraphics[clip=true,scale=0.4]{fig7a.eps}
\includegraphics[clip=true,scale=0.4]{fig7b.eps}
\includegraphics[clip=true,scale=0.4]{fig7c.eps}
\caption{\label{fig_gap}Band gap as a quadratic function of mixing content for (a) MA(Sn$_{1-x}$Pb$_x$)I$_3$ from experiment~\cite{Hao14jacs}, where phase transition occurs from the pseudocubic $P4mm$ phase for MASnI$_3$ to the tetragonal $I4cm$ phase for MAPI, (b) MAPb(I$_{1-x}$Br$_x$)$_3$, reprinted figure with permission from~\cite{yucj16prb}, Copyright 2016 by the American Physical Society, and (c) MAPb(I$_{1-x}$Cl$_x$)$_3$, reprinted figure with permission from~\cite{yucj17jps}, Copyright 2017 by the Elsevier B.V., in pseudocubic phases with DFT calculations.}
\end{center}
\end{figure}

Substitution or mixing on the B site can directly alter the conduction band dominated by B $p$ states without the severe change of crystal structure. For example, replacing Pb with isovalent Sn in MAPI reduces the band gap by $\sim$0.3 eV due to a slight downward shift of CBM by weaker Sn 5$p$ states than Pb 6$p$ states~\cite{Hao14jacs}. It also induces a phase transition from the tetragonal $I4cm$ phase for MAPI to the pseudocubic $P4mm$ phase for MASnI$_3$, being beneficial to the light absorber. However, Sn$^{2+}$ is less chemically stable in the octahedral network due to its ease of oxidation to Sn$^{4+}$, resulting in the degradation of PSCs. Another isovalent Ge$^{2+}$ is further unstable owing to its lower binding energy of 4$p$ electrons. When mixing Pb with Sn to form solid solutions MA(Sn$_{1-x}$Pb$_x$)I$_3$, the band gap was found to follow the quadratic function of mixing content $x$ (see Fig.~\ref{fig_gap}), occurring a band inversion associated with a systematic change in the atomic orbital composition of the conduction and valence bands~\cite{Hao14jacs,Mosconi15jmca,Walsh05jpcb}.

Since the X-site anion dominates the valence bands, its substitution can be expected to change the VBM~\cite{Lindblad15jpcc}. As going from I to Br to Cl, the valence band composition changes from $5p$ to $4p$ to $3p$, resulting in a monotonic increase in electron binding energy (higher ionization potential). Accordingly, the band gap increases from 1.5 eV for MAPI to 2.10 eV on Br~\cite{yucj16prb} to 2.70 eV on Cl substitution~\cite{yucj17jps}. For the case of FAPbX$_3$, the substitution of Br for I increases the band gap from 1.48 eV to 2.23 eV as found in experiment~\cite{Eperon14ees}. For the former case of MA(Sn$_{1-x}$Pb$_x$)I$_3$, mixing I with Br or Cl can give band gaps changing according to the quadratic function of mixing content (Fig.~\ref{fig_gap}),
\begin{equation}
E_g(x)=E_g(0)+[E_g(1)-E_g(0)-b]x+bx^2 \label{eq_bandgap}
\end{equation}
where $b$ is the bowing parameter reflecting the fluctuation degree in the crystal field and the nonlinear effect from the anisotropic binding~\cite{yucj16prb,yucj17jps,Noh13nl}. If the bowing parameter would be zero, the band gap also could follow the Vegard's law like lattice constant. Comparing between MAPb(I$_{1-x}$Br$_x$)$_3$ and MAPb(I$_{1-x}$Cl$_x$)$_3$, they are 0.185 eV (0.33 eV in experiment~\cite{Noh13nl}) and 0.873 eV respectively, indicating the larger compositional disorder and low miscibility for the latter case (Fig.~\ref{fig_gap}(b) and (c)).

\subsection{Transportation of charge carriers}
In the halide perovskites, the major charge carriers are conduction electrons and holes created by several external factors. They can couple each other by an electrostatic interaction to form a kind of quasiparticle, namely exciton. Therefore, effective masses of these charge carriers and exciton binding energy are of importance in checking the suitability for charge transport layer in solar cells.

Effective masses of conduction electrons and holes can be calculated from the lower conduction band and upper valence band near the R point for the cubic phase, which can be approximated as a parabolic function of momentum $k$ as follows.
\begin{equation}
E=\frac{\hbar^2}{2m^*}k^2 \Longrightarrow m^*=\hbar^2\left[\frac{\partial^2E}{\partial k^2}\right]^{-1}
\end{equation}
This parabolic approximation can be slightly broken when considering the SOC effect in the hybrid halide perovskites as discussed above. Therefore, the calculated effective masses depend somewhat on the theoretical method and the complexity of band dispersion. The $GW$+SOC calculation yielded average values of $m_e^*/m_e=0.19$ (0.28) and $m_h^*/m_e=0.25$ (0.13) for cubic MAPI (MASnI$_3$)~\cite{Umari14sr}, whereas PBE with an inclusion of van der Waals (vdW) correction~\cite{Grimme06jcc} can give similar values of $m_e^*/m_e=0.20$ (0.34) and $m_h^*/m_e=0.23$ (0.43) for MAPI (MAPbCl$_3$)~\cite{WGeng14jpcc,yucj17jps}. The electron effective masses are larger than those of GaAs (0.07) and CdTe (0.11), whereas the holes have distinctly lower masses than those (0.5 and 0.35, respectively), predicting a high hole mobility in the hybrid halide perovskites. For the cubic CsSnX$_3$, the hole effective masses were predicted to decrease going from Cl to Br to I by QS$GW$ calculations~\cite{Huang13prb}. Since the mobility is inversely proportional to the effective mass, the halide perovskites are expected to exhibit ambipolar and large diffusion lengths for electrons and holes~\cite{Giorgi13jpcl}.

An exciton can be viewed as hydrogen atom with differences attributed to (i) a replacement of the electron and nucleus masses by the effective masses and (ii) an introduction of a dielectric constant $\varepsilon$ of material. Then, the exciton binding energy can be obtained using the effective masses and dielectric constant in the following equation,
\begin{equation}
 \label{eig_exciton}
E_b=\frac{m_ee^4}{2(4\pi\varepsilon_0)^2\hbar^2}\frac{m_r^*}{m_e}\frac{1}{\varepsilon^2}\approx13.56\frac{m_r^*}{m_e}\frac{1}{\varepsilon^2}~(\mbox{eV})
\end{equation}%
where $m_r^*$ ($1/m_r^*=1/m_e^*+m_h^*$) is the reduced effective mass. The static dielectric constant can be calculated by DFPT method without or with the electron-hole coupling as will be discussed below. If this modelling is valid as the extending radius of the lowest bound state ($a^*=\varepsilon\frac{m_e}{m_r^*}a_B$, $a_B$ the Bohr radius) is larger than the lattice constant, the exciton is the weak exciton, called as the Mott-Wannier exciton. Otherwise, it is a Frenkel exciton, being extremely localized. For MAPI (MAPbBr$_3$), the exciton binding energy $E_b$ and effective radius $a^*$ were calculated to be 45 (99) meV and 3.0 (1.9) nm~\cite{yucj16prb}, indicating that the excitons are likely to be of the Mott-Wannier type.

Charge carrier transport can be characterized by the diffusion length, defined as the average length of carrier traveling before recombination. This can be calculated using the diffusivity $D$ and charge carrier lifetime $\tau$ by $L_d=\sqrt{D\tau}$. The diffusivity can be directly related to the mobility ($\mu$) by the Einstein relation $\mu=Dq/k_BT$ due to the fact that transport is limited by carrier scattering and carrier effective mass. The $L_d$ must be long enough so that the charge carriers can reach the contacts in solar cells, as reported to be considerably longer in MAPI than other semiconductors. This can be partly attributed to the defect-tolerance of hybrid halide perovskites~\cite{Li15sr}. Providing effective masses of $<$0.2$m_e$ for MAPI, the carrier mobility was calculated to be modest ($<$100 cm$^2/($V$\cdot$s)) compared to the inorganic semiconductors Si or GaAs ($>$1000 cm$^2/($V$\cdot$s))~\cite{Stranks15nn}. Such limitation of carrier mobility can be caused by strong scattering. Zhao et al.~\cite{Zhao16sr} have found that by applying PBE+SOC approach the electron-acoustic phonon couplings in MAPI are weak, and charge carriers are scattered predominantly by charged defects or impurities.

\subsection{Dielectric constant and photoabsorption coefficient}
The frequency dependent complex dielectric functions can be calculated within DFPT without the electron-hole coupling effects. Although such excitonic effects can be calculated by solving the Bethe-Salpeter equation for the two-body Green's function without or with local field effect~\cite{Rohlfing00prb}, ignoring electron-hole coupling still yields quite a reasonable result in good agreement with experiment for the small band gap semiconductors~\cite{Onida02rmp}. Given a self-consistent QS$GW$ potential, the the polarizability $P(\mathbi{q}, \omega)$ is obtained using the random phase approximation (RPA), and then the inverse dielectric function is obtained from $\varepsilon^{-1}=[1-\nu(\mathbi{q})P(\mathbi{q}, \omega)]^{-1}$ where $\nu(\mathbi{q})$ is the bare Coulomb interaction~\cite{Leguy16ns}. The macroscopic dielectric function $\varepsilon_{\mtxt{M}}$ is given as follows,
\begin{equation}
\varepsilon_{\mtxt{M}}=\lim_{q\rightarrow 0}\frac{1}{\varepsilon^{-1}_{\scbi{G}=0,\scbi{G}'=0}}
\end{equation}
Then, the frequency dependent photoabsorption coefficient is given as follows,
\begin{equation}
\alpha(\omega)=\frac{2\omega}{c}\sqrt{\frac{\left[\mbox{Re}^2\varepsilon_{\mtxt{M}}(\omega)+\mbox{Im}^2\varepsilon_{\mtxt{M}}(\omega)\right]^{1/2}-\mbox{Re}\varepsilon_{\mtxt{M}}(\omega)}{2}} 
\end{equation}

\begin{figure}[!th]
\begin{center}
\includegraphics[clip=true,scale=0.45]{fig8a.eps}
~~~~\includegraphics[clip=true,scale=0.45]{fig8b.eps} \\ \vspace{7pt}
\includegraphics[clip=true,scale=0.15]{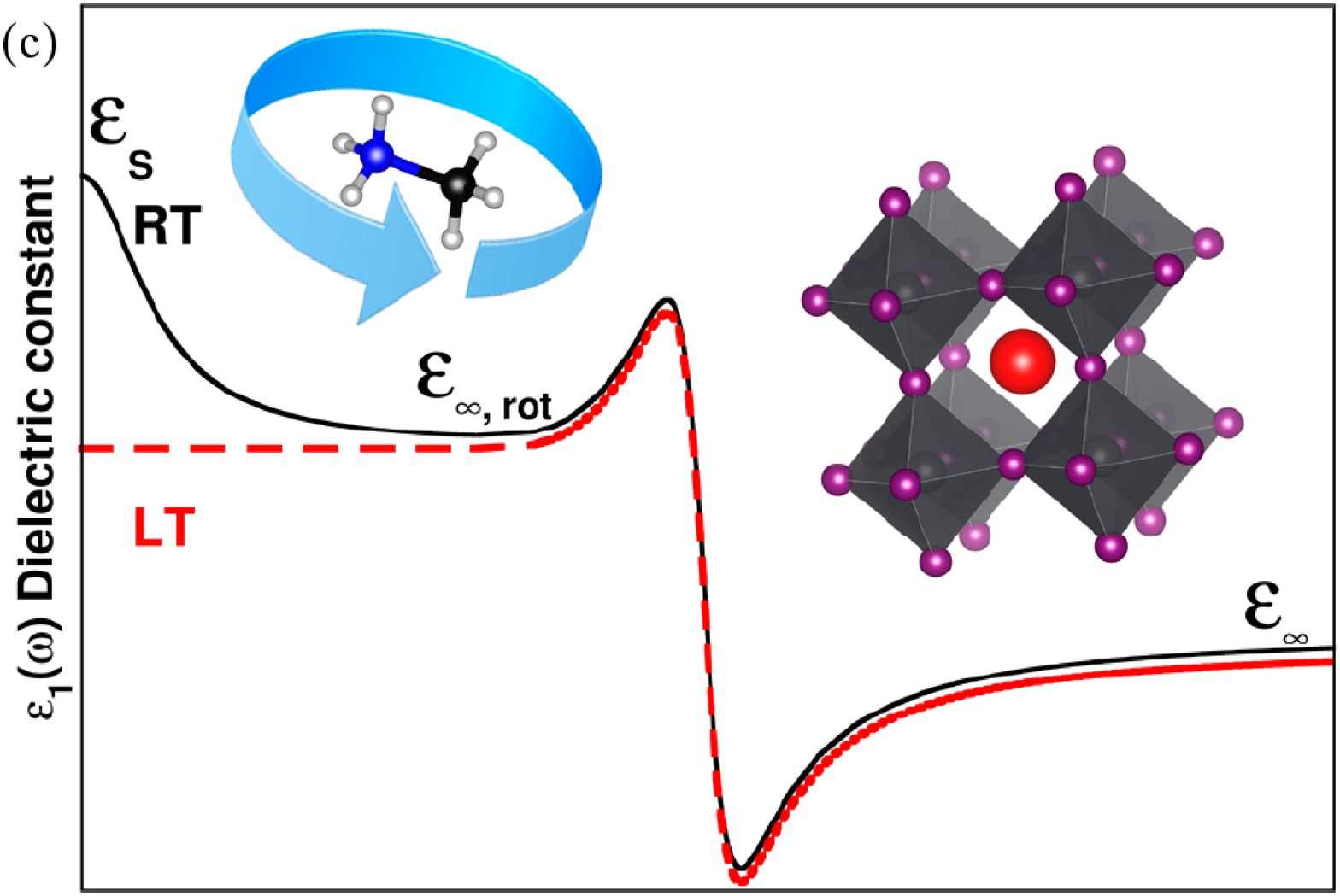}
\includegraphics[clip=true,scale=0.3]{fig8d.eps}
\caption{\label{fig_opt}(a) Real and (b) imaginary part of macroscopic dielectric function, and (d) photoabsorption coefficient as increasing the Cl content in MAPb(I$_{1-x}$Cl$_x$)$_3$. Reprinted figure with permission from~\cite{yucj17jps}, Copyright 2017 by the Elsevier B.V. (c) Schematic view of dielectric function in relation to MA$^+$ rotations and vibration phonons in the low temperature (LT) orthorhombic (red dash line) and room temperature (RT) tetragonal (black solid line) phases of MAPI. Reprinted figure with permission from~\cite{Even14jpcc}, Copyright 2014 by the American Chemical Society.}
\end{center}
\end{figure}

Figure~\ref{fig_opt} shows the macroscopic dielectric functions and the photoabsorption coefficients as functions of photon energy in hybrid halide mixed perovskites MAPb(I$_{1-x}$Cl$_x$)$_3$, calculated using PBE+vdW-D2 functional within VCA as the mixing content $x$ increases~\cite{yucj17jps}. The real part is related to polarizability of the medium in response to an oscillating electric field. For the hybrid solid solutions MAPb(I$_{1-x}$Br$_x$)$_3$ and MAPb(I$_{1-x}$Cl$_x$)$_3$, the static dielectric constants decrease as the mixing content $x$ increases~\cite{yucj16prb,yucj17jps}. The rotation of molecular cation MA$^+$ is of quite importance in the explanation of these variation tendencies. In fact, the static dielectric constant, $\varepsilon_s=\lim_{\omega\rightarrow 0}\mbox{Re}\varepsilon_{\mtxt{M}}(\omega)$, is dominated by rotational motion of MA molecule that has an intramolecular dipole, while the high-frequency or optical dielectric constant, $\varepsilon_\infty\approx\varepsilon_{\mtxt{M}}(\omega\sim10^{15}\mbox{ Hz})$ is related to the vibrational polar phonons of the lattice (see Fig.~\ref{fig_opt}(c))~\cite{Even14jpcc}. The MA molecular dipoles screen the Coulombic interaction between photo-excited electrons and holes in the Pb-I sublattice~\cite{Hakamata16sr,FZheng15jpcl,Grancini15np}, leading to a reduction of exciton binding. The MA cation is situated in a cuboctahedral cage, and thus, when going from I to Br to Cl, the size of cage decreases, inducing a restriction of molecular motion and a decrease of static dielectric constant with a reduction of exciton screening. In accordance to these tendency, the photoabsorption onset and the first peak shift to a higher photon energy, i.e. shorter wavelength light, as increasing the mixing content. There exist other first-principles calculations of optical properties for CsSnI$_3$~\cite{GSong17ra}, CsPbX$_3$~\cite{He18jpcl}, MASnX$_3$~\cite{Feng14jpcc} and MABaI$_3$~\cite{Kumar16prb}.

\section{Phonon and material stability}
Crystalline materials in general exhibit phase transition upon changing temperature and pressure. A certain phase with a specified crystal structure can exist in stable state in the certain range of temperature and pressure, and at the critical value, changes into another phase. Such stability at the thermodynamic condition is called phase stability. On the other hand, crystalline materials can be chemically decomposed into several components, which may be in crystalline and/or gaseous states. This is called chemical stability. During such chemical decomposition, heat can be generated (exothermic reaction) or absorbed (endothermic reaction). For the former case, the chemical decomposition can occur spontaneously, being intrinsic chemical instability that can be considered at different thermodynamic conditions, while for the latter case the reaction can be triggered and progressed with extrinsic factors such as moisture, light and heating, being extrinsic chemical instability. Both the phase stability and intrinsic chemical instability at the certain thermodynamic condition can be predicted by performing phonon calculations to determine the entropy~\cite{Butler16cs}.

The Gibbs free energy of a compound as a function of temperature $T$ and pressure $P$ is given as
\begin{equation}
G(T,~P)=F(T,~V)+PV 
\end{equation}
where $F(T,~V)$ is the Helmholtz free energy as a function of temperature and volume $V$. For the $PV$ term, the total energies are calculated as varying the unit cell volume, fitted into the empirical equation of state for solid~\cite{Birch47pr} to obtain $E(V)$ function, and then the pressure is estimated by conducting a differentiation like $P=-(\partial E/\partial V)_T$~\cite{yucj14pb}. Within the adiabatic approximation, $F(T,~V)$ can be separated into ionic vibrational and electronic contributions~\cite{Grabowski07prb,Baroni01rmp},
\begin{equation}
\label{eq_Free}
F(T,~V) = F_{\mtxt{vib}}(T,~V) + F_{\mtxt{el}}(T,~V)\simeq F_{\mtxt{vib}}(T,~V) + E(T=0~\mbox{K},~V)
\end{equation}
In the electronic Helmholtz free energy, $F_{\mtxt{el}}(T,~V)=E(T=0~\mbox{K},~V)-TS_{\mtxt{el}}$, the $TS_{\mtxt{el}}$ term is ignored because the electronic temperature effect is negligible for non-metallic systems at the room temperature vicinity~\cite{Landau1960}. Within the quasiharmonic approximation (QHA), the ionic term $F_{\mtxt{vib}}$ can be calculated as follows~\cite{yucj14pb,Grabowski07prb},
\begin{equation}
\label{equ_F_vib}
F_{\mtxt{vib}}(T,~V)=3Mk_BT\int_0^{\omega_L}\ln\left\{2\sinh\left[\frac{\hbar\omega(V)}{2k_BT}\right]\right\}g(\omega)d\omega
\end{equation}
where $\omega(V)$ is the phonon frequency as a function of volume, $M$ the atomic mass, $g(\omega)$ the normalized phonon DOS and $\omega_L$ the maximum of the phonon frequencies. The phase stability or chemical intrinsic stability can be estimated by the Gibbs free energy difference between the two phases or the products and the reactants.

\subsection{Phonon dispersion and phase stability}
It is crucial to get a precise insight of phonon dispersions in the halide perovskites for understanding material processes such as ionic transport, and recombination and scattering of charge carriers, as well as material stabilities. The phonon dispersion is computationally accessible via lattice dynamics calculations, in which the dynamic matrix is constructed via the Hessian matrix as the second derivatives of the total energy with respect to the atomic displacements that can be obtained directly from DFT calculations. The Hessian matrix (or interatomic force constants: IFCs) can be calculated either in real space by performing force calculations on a series of symmetry-inequivalent displaced structures with an use of supercell~\cite{Parlinski97prl,Chaput11prb} or in reciprocal space through perturbation theory (e.g. DFPT~\cite{Baroni01rmp}). Diagonalizing the dynamic matrix yields a set of eigenvectors (phonon modes) and eigenvalues (phonon frequencies)~\cite{Butler16cs}. The lattice dynamics in DFT is one of the highest expensive calculations, and thus, classical or semi-classical MD based on interatomic potential function~\cite{Catlow88pma} can be used to overcome the limitations of DFT. 

For the hybrid perovskite MAPI in the cubic, tetragonal and orthorhombic phases, the phonon bands were calculated in the harmonic approximation~\cite{Quarti14jpcl,Brivio15prb}. Confirming that the calculated phonon spectra were in quite well agreement with the experimental data, these calculations indicate that the lowest bands below 100 cm$^{-1}$ are assigned to the bending and stretching of the Pb$-$I bonds, i.e. diagnostic modes of the inorganic cage, the medium bands in the range from 100 to 150 cm$^{-1}$ are the coupled modes to the molecular motion, and the highest frequency branches in the range from 300 to 3300 cm$^{-1}$ are associated with the vibration and bond stretching of MA cation. For other hybrid halide perovskites MAPbX$_3$ (X = Br, Cl), similar arguments were found to be valid, putting special emphasis on the reorientation of the MA cations, which plays a key role in shaping the vibrational spectra of the different compounds~\cite{Leguy15nc,Chen15pccp,Leguy16pccp,Even16n}. It was found that MAPI exhibits double-well instabilities at the Brillouin zone boundary and remarkably short phonon quasiparticle lifetime, being associated with low thermal conductivity, and the optical phonon scattering is stronger than the acoustic one at room temperature in these materials~\cite{Beecher16ael,Leguy16pccp,Whalley16prb}. Vibrational entropy was proved to play a crucial role in determining the stable phase, e.g. between the pseudocubic and tetragonal phase for MAPI, such that the materials favour structures that maximize the number of soft intermolecular interactions as the temperature rises~\cite{Butler16prb}.

\begin{figure}[!th]
\begin{center}
\includegraphics[clip=true,scale=0.2]{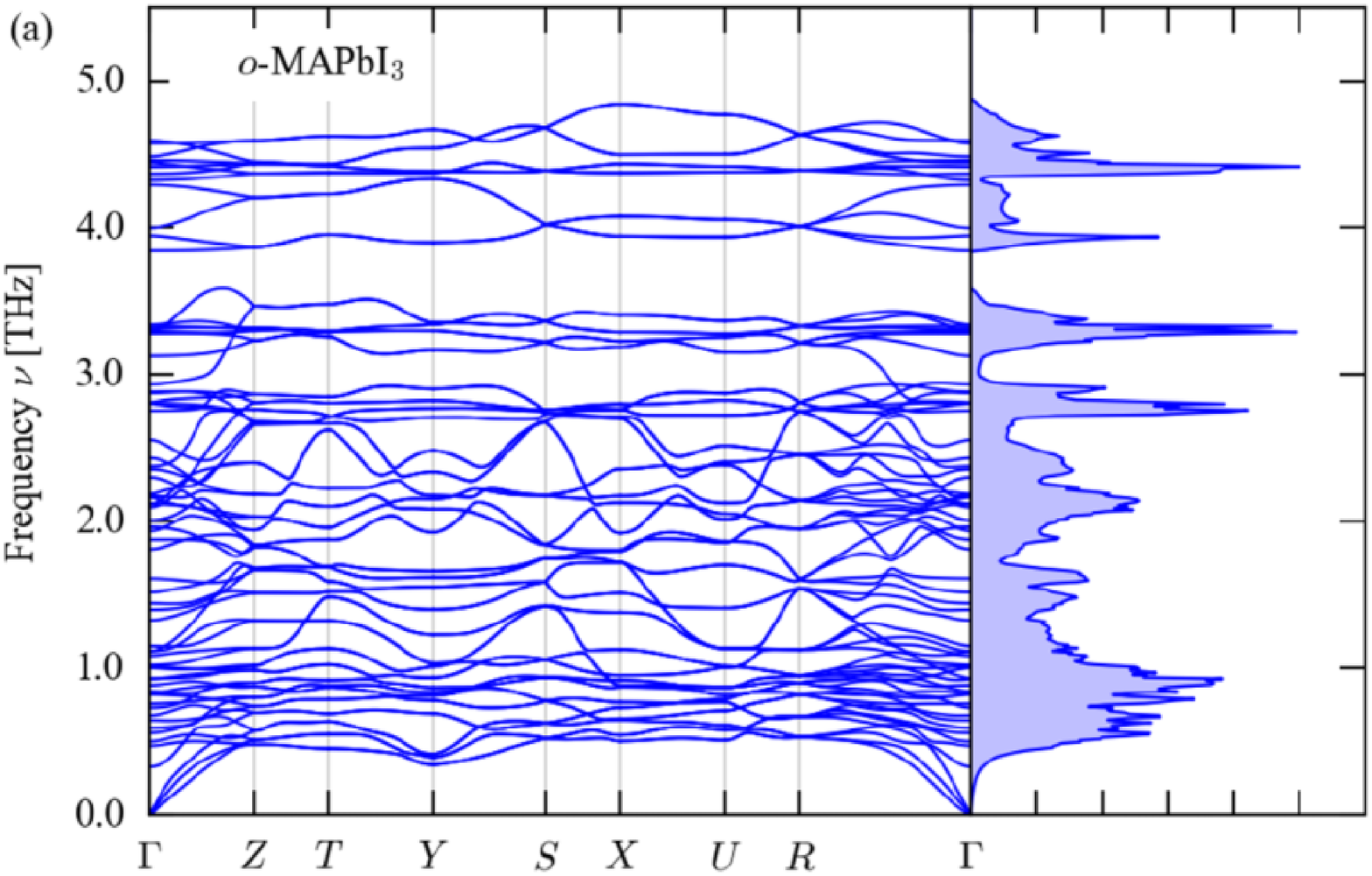}
\includegraphics[clip=true,scale=0.2]{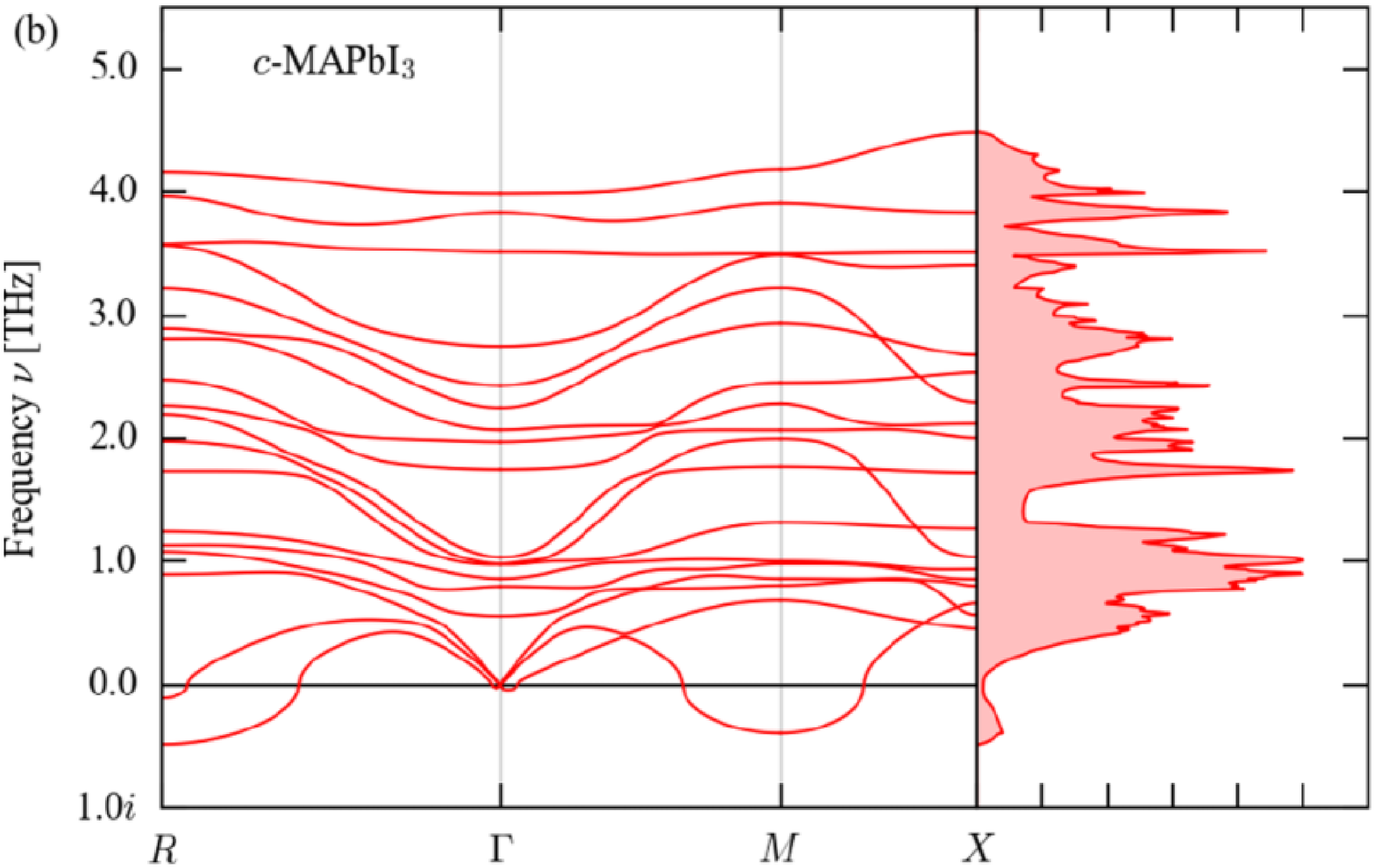} \\ \vspace{7pt}
\includegraphics[clip=true,scale=0.35]{fig9c.eps}
~~~~~\includegraphics[clip=true,scale=0.2]{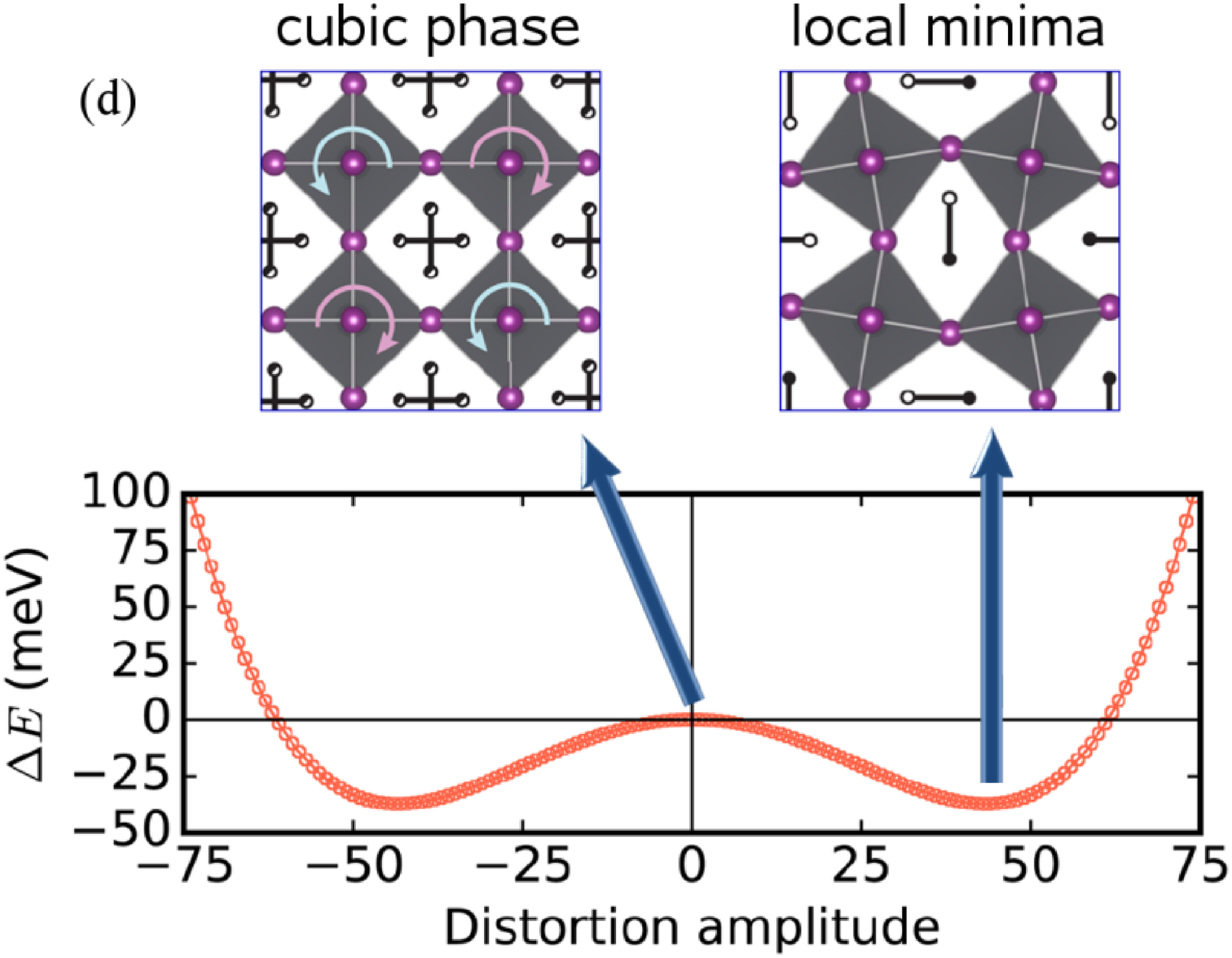}
\caption{\label{fig_phonon}Phonon dispersions and density of states for (a) orthorhombic and (b) cubic MAPbI$_3$,
reprinted figure with permission from~\cite{Whalley16prb}, Copyright 2016 by the American Physical Society, (c) cubic CsPbI$_3$, adapted from~\cite{jong18jmca} with permission of The Royal Society of Chemisty, and (d) double-well potential with a saddle point corresponding to the cubic phase and two local minima, adapted from~\cite{Beecher16ael} with permission of the American Chemical Society.}
\end{center}
\end{figure}
It is important to understand the anharmonic effects in the phase transition of halide perovskites. When the potential energy of a crystalline solid is expanded as a Taylor series of ionic displacements, only the second term ($d^2U/dr^2$) is considered in the harmonic approximation with the temperature-independent frequencies and infinite lifetime, while the effects of temperature and first-order anharmonicity can be considered in QHA~\cite{Buckeridge13prb}. It is associated with the imaginary (or negative) frequency, i.e. soft mode. While all the phonon modes have positive frequencies in the tetragonal and orthorhombic phases (Fig.~\ref{fig_phonon}(a)), there are two imaginary frequency acoustic modes in the cubic phase, of which centres are around the $R[\frac{2\pi}{a}(\frac1{2},\frac1{2},\frac1{2})]$ and $M[\frac{2\pi}{a}(\frac1{2},\frac1{2},0)]$ points, corresponding to the $<$111$>$ and $<$110$>$ directions (Fig.~\ref{fig_phonon}(b)). As a common feature of the perovskites, these soft modes indicate a double-well characteristic in the potential energy surface, and thus a dynamic instability of the cubic structure as observed in inelastic X-ray scattering measurements~\cite{Beecher16ael}. Such instabilities represent antiferroelectric distortions associated with collective rotation and tilting of the corner-sharing octahedral framework that can be observed directly in MD~\cite{Frost14aplm,Quarti16ees}. In the double-well potential, the cubic phase is a saddle point between two equivalent broken-symmetry phases (Fig.~\ref{fig_phonon}(d)), and the barriers are 37 and 19 meV for the $R$ and $M$ modes, being comparable to $k_BT$ at room temperature~\cite{Whalley16prb}. The phase transitions from cubic structure can be understood as a condensation of these soft modes, as proved by solving the time-dependent Kohn-Sham equation describing the nuclear motion in the double-well potential~\cite{Whalley16prb,Adams16jpcm,Skelton16prl}.

The soft phonon modes are also found in the all-inorganic halide perovskites. For the case of CsSnI$_3$ perovskite, the temperature dependent lattice dynamics were carried out for the four different phases within QHA, revealing the strong anharmonic effects such as soft modes~\cite{Silva15prb,Patrick15prb}. However, the calculated phase transition temperatures fail in quantitative agreement with experimental values, requiring further studies to eliminate these discrepancies. Yang et al.~\cite{Yang17jpcl} systematically investigated the phonon dispersions of CsSnX$_3$ and CsPbX$_3$ (X = F, Cl, Br, I) perovskites, verifying the phase instabilities of cubic structures related to the spontaneous octahedral tilting. In all the eight cases, double-well potentials as a function of distortion amplitude (Q), given by
\begin{equation}
E(Q)=aQ^2+bQ^4+O(Q^6)
\end{equation}
with fitting parameters $a$ and $b$, were found for soft phonon modes with barrier heights ranging from 108 to 512 meV, assessing the chemical and thermodynamic driving forces for these instabilities. When compared with the hybrid halide perovskites, the soft mode is found surprisingly even at the zone centre point $\Gamma$ (Fig.~\ref{fig_phonon}(c)), which is related to the ferroelectric distortion. Marronnier et al.~\cite{Marronnier17jpcl,Marronnier18an} deeply investigated this anomaly at $\Gamma$ for the cubic and tetragonal phases of CsPbI$_3$. As a polar mode, the soft phonon mode found at $\Gamma$ for cubic CsPbI$_3$ is linked to the displacements of Cs$^+$ cation in one direction and of I$^-$ anion in the opposite direction. In spite of such displacements, ferroelectricity has not been observed at a macroscopic scale due to the oscillations along this polar soft mode. Volume relaxation with tight convergence thresholds (10$^{-4}$ Ry/bohr for the force and 10$^{-14}$ for the phonon self-consistency) and frozen phonon calculations remove the soft modes at $\Gamma$. Based on those results, it was concluded that the strongly anharmonic mode at $\Gamma$ will not condensate at lower temperatures, whereas the remaining phonon instabilities at the edge points of $M$ and $R$ are responsible for soft modes that condensate at lower temperatures to induce the phase transition~\cite{Marronnier17jpcl}.

\subsection{Thermodynamic miscibility in solid solutions}
Forming solid solutions or alloys by mixing equivalent two or three elements at the same site can provide a direct and easy route to improve the performance of materials. As such, perovskite solid solutions were found to have a controlled band gap and improved stability; in lead-based hybrid halide perovskites APbX$_3$, the best efficiency can be obtained by mixing of organic cations (MA/FA) on the A-site and halides on the X site. In such mixing cases, it is important to understand whether the solid solution is stable against phase separation in the entire range of composition, where configurational entropy can play an important role in describing the disorder of site occupancy~\cite{Butler16cs,Mosconi13jpcc}. Thermodynamic miscibility for mixing two components, e.g. ($1-x$)ABX$_3+x$A$^{\prime}$BX$^{\prime}_3$, can be calculated by the Helmholtz free energy difference given as
\begin{equation}
\Delta F(x, T)=\Delta U(x, T)-T\Delta S(x)
\end{equation}
The first term, internal energy difference, is written as follows~\cite{Caetano16prb},
\begin{eqnarray}
\Delta U(x, T)=\frac{\sum_k\Delta U_k(x)g_ke^{-E_k(x)/k_BT}}{\sum_kg_ke^{-E_k(x)/k_BT}} \\
\Delta U_k(x)=E_k(x)-[(1-x)E_{\mtxt{ABX}_3}+xE_{\mtxt{A}^{\prime}\mtxt{BX}^{\prime}_3}]
\end{eqnarray}
where $E_k(x)$ is the DFT total energy of the alloy with the mixing content $x$ and the configuration number $k$, and $g_k$ is the degeneracy representing the number of configurations with the same energy $E_k$. At room temperature, the internal energy of mixing can be well fitted into the subregular solution expression~\cite{Brivio16jpcl,Caetano16prb},
\begin{equation}
\Delta U(x)=\Omega x(1-x),~~\Omega=\alpha+\beta x
\end{equation}
with the fitting parameters, $\alpha$ and $\beta$. Then, the entropy of alloy is given in the ideal solution limit as follows,
\begin{equation}
\Delta S(x)=-k_B[x\ln x+(1-x)\ln(1-x)] 
\end{equation}
which is expected for a random alloy at high temperatures.

\begin{figure}[!th]
\begin{center}
\includegraphics[clip=true,scale=0.18]{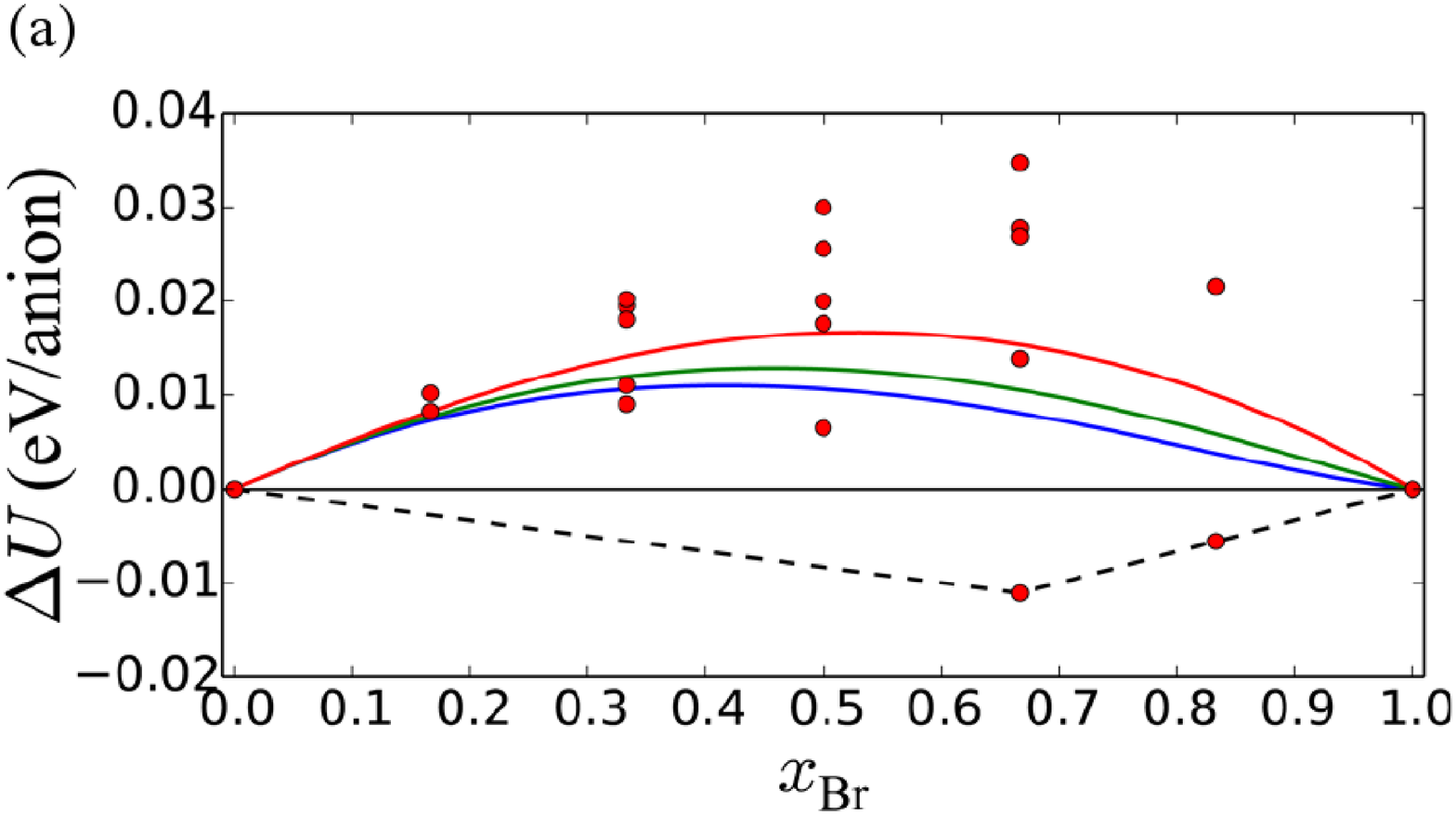}
\includegraphics[clip=true,scale=0.18]{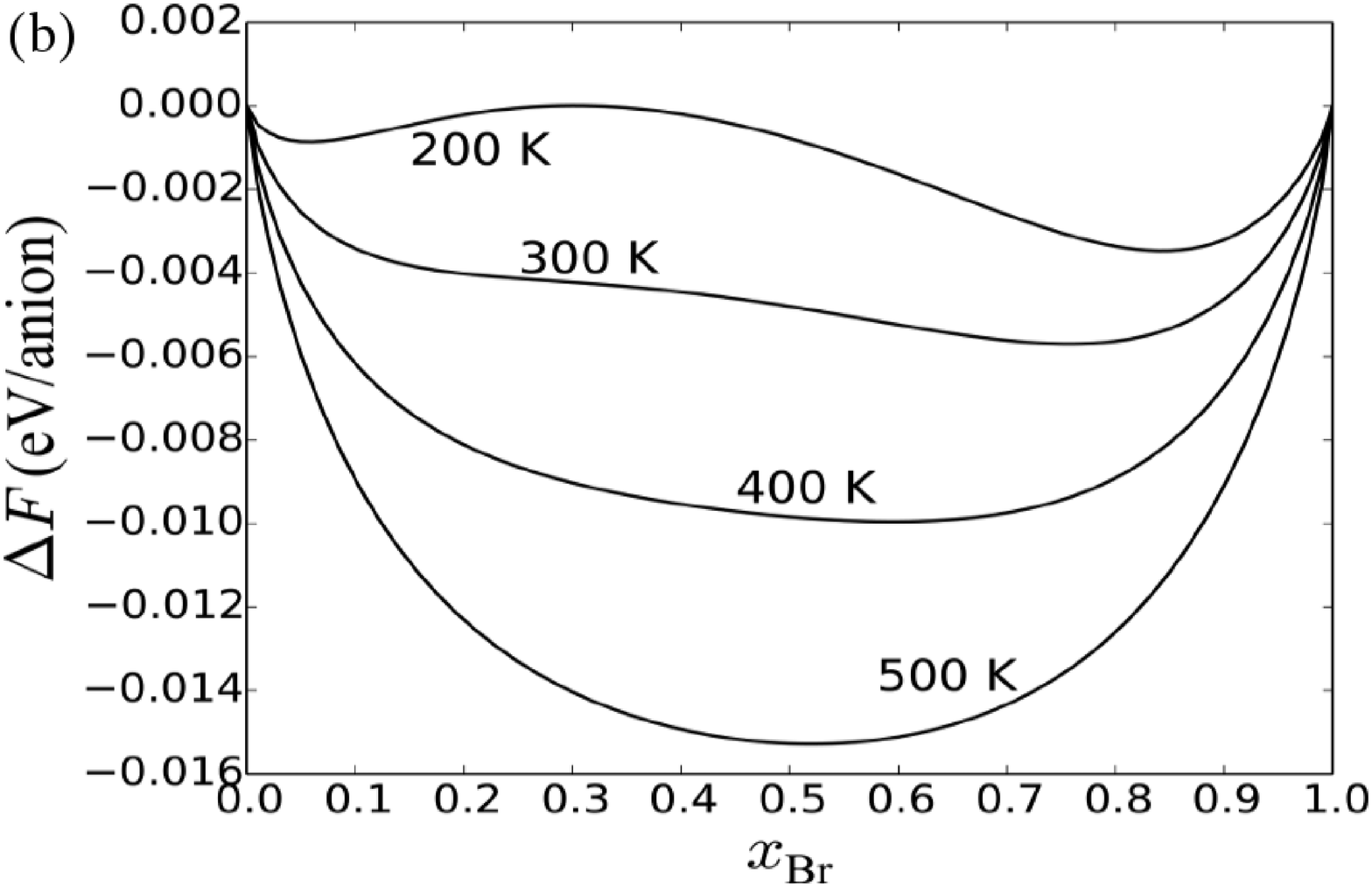} \\
\includegraphics[clip=true,scale=0.2]{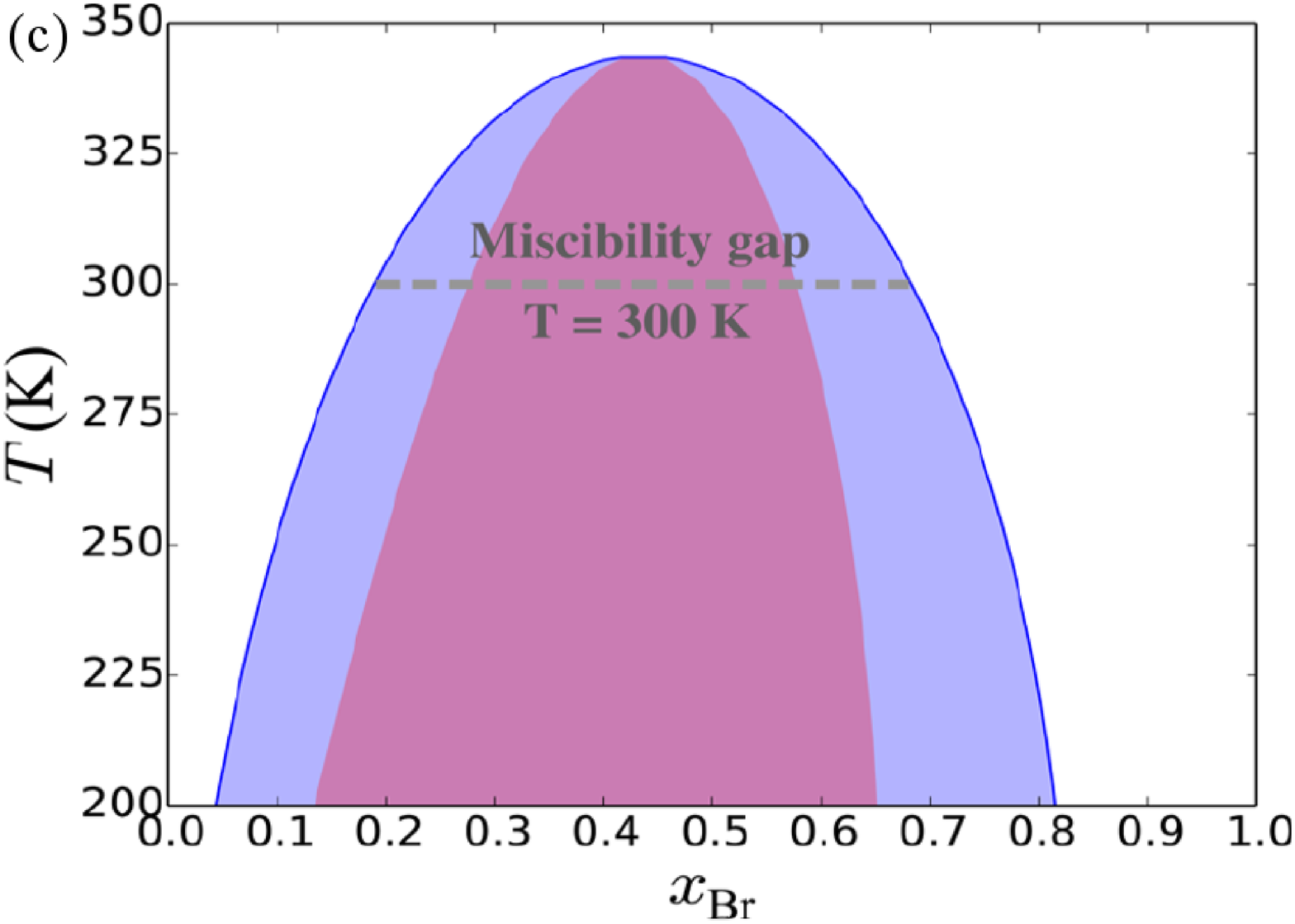}
~~~\includegraphics[clip=true,scale=0.3]{fig10d.eps}
\caption{\label{fig_misc}(a) Internal energy, (b) Helmholtz free energy, and (c) phase diagram of mixing in MAPb(I$_{1-x}$Br$_x$)$_3$. In part (a), the solid lines with blue, green, and red colours refer to 200, 300 K and high T limit, and dashed line represents the convex hull. In part (c), the purple and pink regions are the binodal and spinodal regions. Stable solid solution can be formed in the white region. Adapted from~\cite{Brivio16jpcl} with permission of the American Chemical Society. (d) Helmholtz free energy of Cs$_{1-x}$Rb$_x$PbI$_3$ in cubic phase. Adapted from~\cite{jong18jmca} with permission of The Royal Society of Chemisty.}
\end{center}
\end{figure}

Brivio et al.~\cite{Brivio16jpcl} have provided the thermodynamic insight for photoinstability in the hybrid halide solid solutions MAPb(I$_{1-x}$Br$_x$)$_3$. By generating different configurations (using the SOD code~\cite{Crespo07jpcm}) and applying the generalized quasichemical approximation method~\cite{Sher87prb}, they calculated the internal energy, configurational entropy, Helmholtz free energy of solid solutions and phase diagram as functions of the alloy composition (Fig.~\ref{fig_misc}). At low temperatures, the free energy curve is asymmetric and positive, indicating the existence of a miscibility gap. When increasing the temperature, the curve becomes symmetric and negative since the probability of sampling all possible configurations increases. From the calculated phase diagram, it is clear that there exist a stable region where the solid solution can be formed in stable state against phase separation; e.g. at 300 K, the alloy can not be formed in the region of $0.19<x<0.68$, which is the miscibility gap. Also, the spinodal and binodal points are found, and thus the alloy can present metastable phases in the intervals of $0.19<x<0.28$ and $0.58<x<0.68$ at 300 K. Although it is difficult to directly compare these findings with experiment, some indirect experimental evidences such as blue-shift in optical absorption from I-rich to Br-rich composition were found. For the all-inorganic perovskites such as CsPb(I$_{1-x}$Br$_x$)$_3$~\cite{Yin14jpcl} and Cs$_x$Rb$_{1-x}$SnI$_3$~\cite{Jung17cm}, similar arguments were valid.

\subsection{Chemical stability}
It is well known that the hybrid halide perovskites are weak on external actions such as humidity, ultraviolet light and heat. Also, whether they are stable compounds with respect to the chemical decomposition or not is an important issue. If they are intrinsically stable, the degradation of PSCs can be suppressed by thoroughly protecting the device from the external actions. To check the chemical stability, the following reaction is generally suggested for chemical decomposition of ABX$_3$ perovskites,
\begin{equation}
\mbox{ABX}_3\rightleftharpoons \mbox{BX}_2+\mbox{AX}
\end{equation}
Then, the Gibbs free energy difference between the products and the reactants, $\Delta G=G_{\mtxt{ABX}_3}-(G_{\mtxt{BX}_2}+G_{\mtxt{AX}})$, is an estimation of the chemical stability.

\begin{figure}[!th]
\begin{center}
\includegraphics[clip=true,scale=0.2]{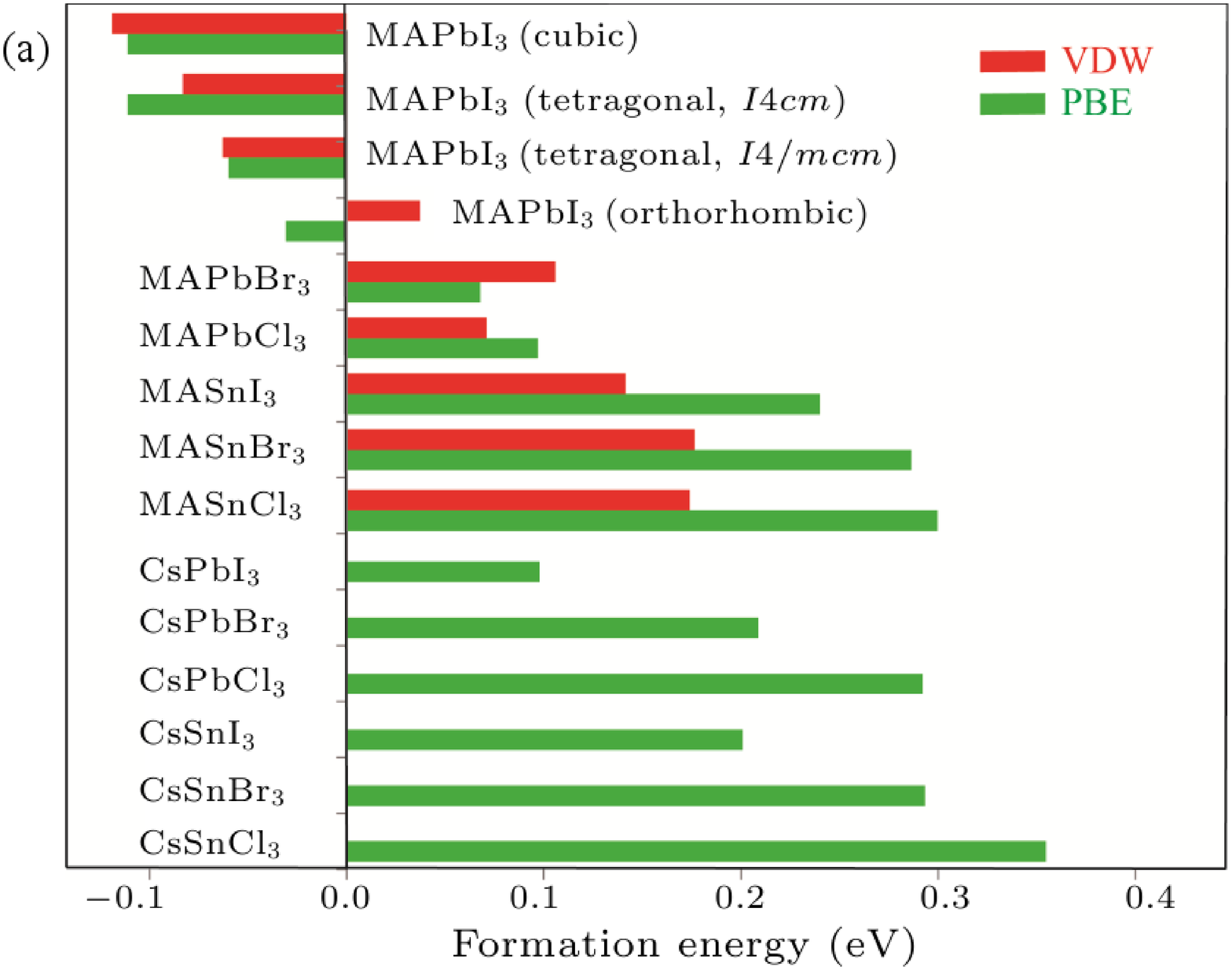}
\includegraphics[clip=true,scale=0.35]{fig11c.eps} \\ \vspace{7pt}
\includegraphics[clip=true,scale=0.35]{fig11b.eps}
~~~~\includegraphics[clip=true,scale=0.3]{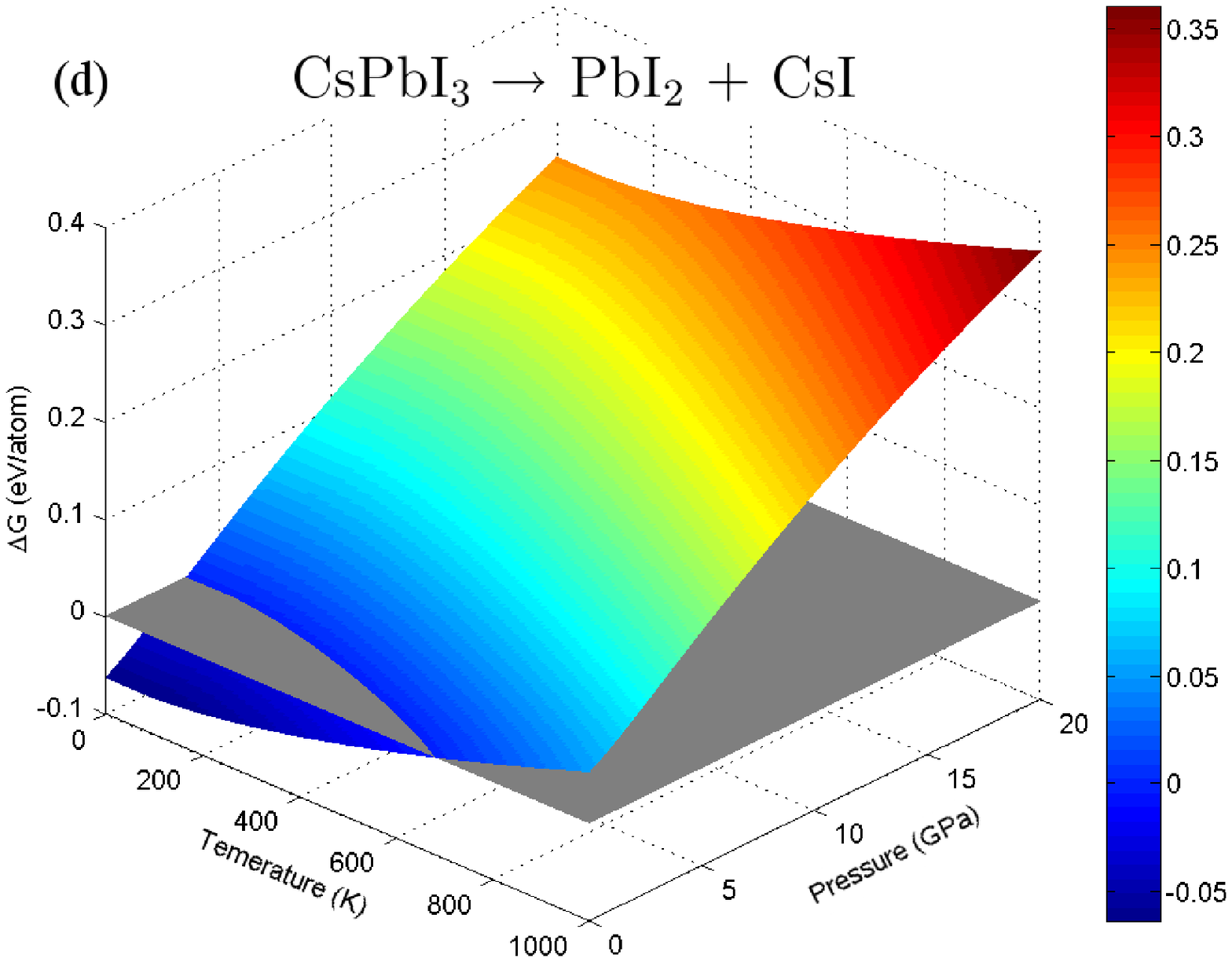}
\caption{\label{fig_chem}(a) Formation energy of halide perovskites ABX$_3$. Reproduced from~\cite{Zhang18cpl}, \copyright 2018 Chinese Physical Society and IOP Publishing Ltd. All rights reserved. (b) Formation energy of solid solution MAPb(I$_{1-x}$Cl$_x$)$_3$. Reprinted with permission from~\cite{yucj17jps}. Copyright 2017 by the Elsevier B.V. (c) Helmholtz free energy difference in Cs$_{1-x}$Rb$_x$PbI$_3$ as varying the mixing content and temperature, and (d) phase diagram of CsPbI$_3$ against chemical decomposition. Adapted from~\cite{jong18jmca} with permission of The Royal Society of Chemisty.}
\end{center}
\end{figure}

At zero temperature and zero pressure condition, simply the DFT total energy difference, i.e. formation energy, can be calculated for this aim~\cite{yucj16prb,yucj17jps,Zhang18cpl}. It was found that for MAPI all the possible phases are unstable for chemical decomposition with PBE functional but including the vdW correction lets the orthorhombic phase stable. As shown in Fig.~\ref{fig_chem}(a), substitution Cs for MA, Sn for Pb, and Br and Cl for I change the formation energy from the negative to positive, indicating the enhancement of intrinsic chemical stability~\cite{Zhang18cpl}. Systematic study on mixing I with Br or Cl in MAPI can provide the turning point of mixing content where the chemical decomposition changes from exothermic to endothermic; 0.2 for MAPb(I$_{1-x}$Br$_x$)$_3$~\cite{yucj16prb} and 0.07 for MAPb(I$_{1-x}$Cl$_x$)$_3$~\cite{yucj17jps} (Fig.~\ref{fig_chem}(b)), which agreed well with the experimental findings. 

At finite temperature, the vibrational contributions were considered by performing the lattice dynamics calculations (mostly via DFPT~\cite{Baroni01rmp}), revealing that interestingly for MAPI and MASI there is no significant change of Gibbs free energy between the reactant and products while for CsSnI$_3$ its free energy decreases faster than that of CsI and SnI$_2$. In other work for MAPI using experimental data for chemical potentials, the free energy difference was calculated to be 0.16 eV/f.u. at 300 K, indicating that the finite temperature effects tend to stabilize the perovskite structure against the spontaneous chemical decomposition under standard thermodynamical condition~\cite{Tenuta16sr}. For mixing Cs with Rb in CsPbI$_3$, the free energy differences tend to increase as the Rb content $x$ increases with a turning point. This turning point decreases as the temperature increases, as shown in Fig.~\ref{fig_chem}(c). The phase diagram for chemical decomposition of CsPbI$_3$ into PbI$_2$ and CsI can be obtained as shown in Fig.~\ref{fig_chem}(d), indicating that CsPbI$_3$ is stable in the temperature range from 0 to 600 K and the pressure range from 0 to 4 GPa~\cite{jong18jmca}, being well fitted with experiment.

\section{Defect formation and ion diffusion}
As in other well-established semiconductors like Si, the understanding of defects in the halide perovskites is of particular interest to material engineers as well as scientists who want to improve the performance of device. In fact, defects in solar cell materials determine a variety of operating processes such as generation and transport of charge carriers (electrons and holes) and ion diffusion, which are highly important for enhancing the efficiency and stability of solar cells. Defects in crystalline solids can be classified into crystallographic defects and impurities, which are in the form of point defects including vacancies, interstitials, antisites, pair-defects (interstitial and vacancies: Frenkel defect; anion and cation vacancies: Schottky defect), substitutional impurity and interstitial impurity, or higher-dimensional defects including dislocations, grain boundaries~\cite{RLong16jacs} and precipitates. Such defects can mediate the ion diffusion or can be mobile themselves inside the solid, which are suggested to be responsible for a current-voltage hysteresis of device and degradation related to moisture and light exposure~\cite{Walsh17nm,Ball16ne}.

Simulations of defect-related phenomena are in highly cost from the viewpoint of memory capacity and computational time. In these simulations, sufficiently large supercells must be adopted to ensure a reliable accuracy of (charged) defect formation energy, because of the errors caused by the finite sizes of supercells~\cite{Kumagai14prb,Kumagai16pra}. To determine the charge transition levels formed by charged defects, the electronic structures of perfect and defect-containing crystals (modelled by supercells) are calculated with a high accuracy, e.g. using hybrid functionals such as HSE06~\cite{HS04jcp} and/or considering SOC effects coupled with many-body theory ($GW$). Formation energies of defects depend on the chemical potential of ingredient elements, which need to consider thermodynamics. The formation enthalpy of a point defect with a charge state $q$ is calculated using the grand canonical expression~\cite{Zhang91prl,Freysoldt14rmp,yucj18jpcl},
\begin{equation}
\Delta H_f[D^q]\cong\{E[D^q]+E_{\mtxt{corr}}[D^q]\}-E_{\mtxt{perf}}-n_i\mu_i+qE_F \label{eq_Hf}
\end{equation}
where each term can be calculated by first-principles method coupled with thermodynamics. On the other hand, modelling and simulation of ion diffusion, which are mostly vacancy mediated, need to suggest preliminary migration paths based on the empirical bond valence sum (BVS) approach~\cite{Brown85acb}, and perform structural optimizations to fix the local configurations, and finally the minimum energy path calculations using the nudged-elastic-band (NEB) approach~\cite{Henkelman00jcp} to determine the energy barrier.

\subsection{Defect formation energy and transition levels}
The most important defects in the halide perovskites ABX$_3$ are the twelve intrinsic point defects: the vacancies (V$_{\mtxt{A}}$, V$_{\mtxt{B}}$, V$_{\mtxt{X}}$), the interstitials (A$_i$, B$_i$, X$_i$), and antisites (A$_{\mtxt{B}}$, A$_{\mtxt{X}}$, B$_{\mtxt{A}}$, B$_{\mtxt{X}}$, X$_{\mtxt{A}}$, X$_{\mtxt{B}}$). For each point defect, various charge states are considered. By using Eq.~\ref{eq_Hf}, formation energies of each point defect with various charge states are calculated as functions of Fermi energy ($E_F$) at different thermodynamic conditions reflecting the growth conditions. In general, equilibrium conditions for the product with its pure constituents (e.g. MAPI is in equilibrium with MA, Pb and I$_2$) give the constraints for the relation between their chemical potentials, resulting in two different scenarios; halide-rich and halide-poor growth conditions. Therefore, two diagrams for defect formation energy can be obtained under these conditions. From these defect formation energy diagrams, it is possible to identify the dominant defect that has the lowest formation energy and thermodynamic charge transition levels.

In the case of tetragonal MAPI, the dominant point defects were found to be the acceptor-type lead vacancy (V$^{-2}_{\mtxt{Pb}}$) under I-rich condition~\cite{yucj18jpcl,Yin14apl,Buin15cm}, while donor-type MA$^{+1}_i$ under I-poor condition~\cite{Yin14apl}. Buin et al.~\cite{Buin14nl} have got a little different findings, such that the major acceptor defects are V$^{-2}_{\mtxt{Pb}}$, V$^{-1}_{\mtxt{MA}}$ and I$^{-1}_i$, whereas donor defects are V$^{+1}_{\mtxt{I}}$ and Pb$^{+2}_i$. Among these, defects V$^{+1}_{\mtxt{I}}$, MA$^{+1}_i$, and V$^{-1}_{\mtxt{MA}}$ were found to possess the lowest formation energies over the entire band gap. The low formation energy of V$_{\mtxt{Pb}}$ is related to the energetically unfavourable $s$-$p$ antibonding coupling~\cite{Yin14apl}. In the case of MAPbBr$_3$ and MAPbCl$_3$ with the cubic phases, Pb$^{-3}_{\mtxt{Br(Cl)}}$ has somewhat comparable formation energy with V$_{\mtxt{Pb}}$~\cite{Buin15cm,Shi15apl}. Interestingly, the dominant defects that have the low formation energies exhibit shallow transition levels in the band gap, while the defects that have higher formation energy (and thus difficult to form) would have deep-trap levels. In addition, all vacancies yield shallow traps and resonances within the band, implying that charge carriers can still move easily to VBM and CBM~\cite{Buin14nl}. It has been reported that only interstitial iodine defect I$_i$ is a deep trap and non-radiative recombination centre for hole~\cite{Du14jmca,WLi17ael}, pointing out the strong dependence of transition levels on the selection of XC functional (HSE+SOC should be used)~\cite{Du15jpcl}. Therefore, it can be suggested that point defects in the hybrid halide perovskites should not contribute to the density of deep traps (recombination centres) that directly controls the diffusion length of charge carriers. However, this can be changed according to the growth condition. For example, the formation energy of deep-level antisite Pb$_{\mtxt{I}}$ may be low enough to significantly contribute to the density of recombination centres under I-rich condition~\cite{Buin15cm,Buin14nl}. In this sense, the optimal growth conditions should be halide-poor for MAPI, and halide-rich for MAPbBr$_3$ and MAPbCl$_3$~\cite{Buin15cm}.

\begin{figure}[!th]
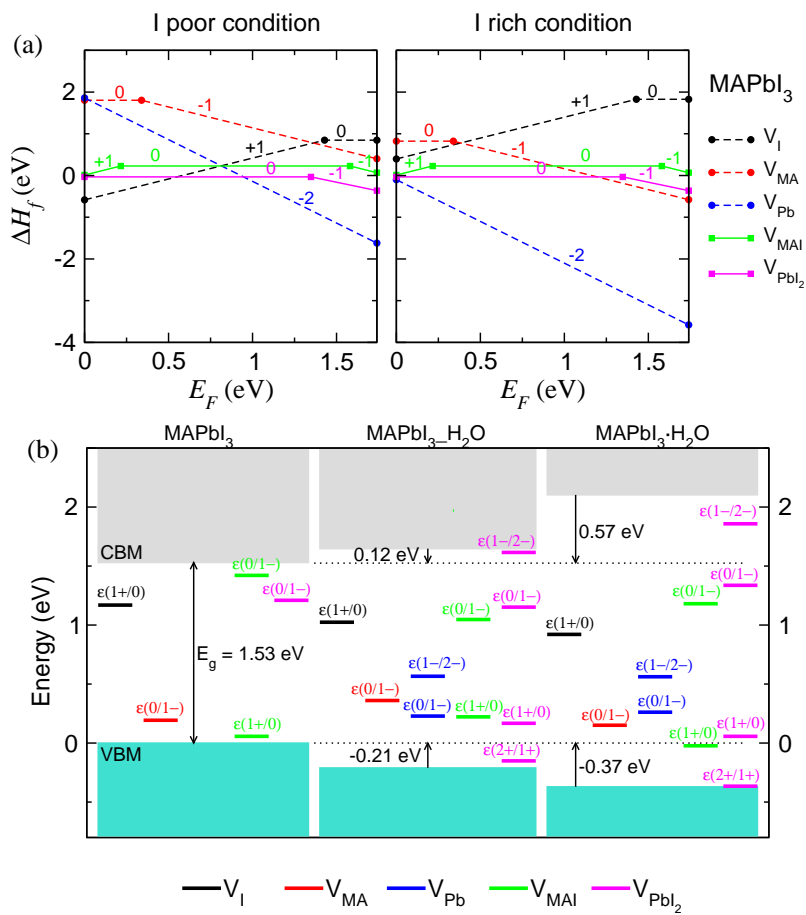

\begin{center}
~\includegraphics[clip=true,scale=0.6]{fig12a.eps} \\ \vspace{3pt}
\includegraphics[clip=true,scale=0.6]{fig12b.eps} 
\caption{\label{fig_def}(a) Formation energy diagrams of vacancy point and pair defects in MAPI under I-poor and I-rich conditions, and (b) band alignment and thermodynamic transition levels in MAPI, water-intercalated MAPI and mono-hydrated MAPI. Reprinted with permission from~\cite{yucj18jpcl}. Copyright 2018 American Chemical Society.}
\end{center}
\end{figure}

The experimentally observed unintentional doping for both n-type and p-type MAPI without any added dopant can be explained by creation of neutral vacancy pair defects, i.e. Schottky defects and Frenkel defects. It was found that Schottky defects such as V$_{\mtxt{PbI}_2}$ and V$_{\mtxt{MAI}}$, which may dominate the defect formation in the halide perovskites under stoichiometric growth conditions, do not make a trap state~\cite{Walsh15acie}. The elemental defects derived from Frenkel defects play the role of unintentional doping sources, implying that n-type and p-type doping can be controlled by carefully choosing the proper atomic composition in the growth procedure~\cite{Kim14jpcl}. Such vacancy pair defects were also considered in the water-intercalated and mono-hydrated MAPI in order to reveal the mechanism of PSC degradation upon exposure to moisture~\cite{yucj18jpcl}. From the calculated binding energy of the vacancy pair defects, it was concluded that the formation of V$_{\mtxt{PbI}_2}$ from the vacancy point defects V$^{-1}_{\mtxt{I}}$ and V$^{+2}_{\mtxt{Pb}}$ is spontaneous in these three compounds, while the formation of V$_{\mtxt{MAI}}$ is less favourable than the formation of individual vacancies V$^{-1}_{\mtxt{I}}$ and V$^{+1}_{\mtxt{MA}}$ in the hydrous compounds. Moreover, all these Schottky defects exhibit deep trap levels in the hydrous compounds (Fig.~\ref{fig_def}), giving an evidence of degradation of PSCs under humid condition~\cite{yucj18jpcl}.

Defect calculations have been performed for other halide perovskites. By using $GW$+SOC method, Li et al.~\cite{Li17apl} investigated the defect physics of cubic CsPbI$_3$. They found that under the Pb-rich conditions the vacancy point defects V$_{\mtxt{Pb}}$ and V$_{\mtxt{I}}$ are the dominant acceptor and donor defects and can pin the Fermi energy in the middle of the band gap due to their comparable formation energies. Under the Pb-poor conditions, the acceptor V$_{\mtxt{Pb}}$ has a high density and thus can generate a high density of hole carriers. While the antisite Pb$_{\mtxt{Cs}}$ acts as the recombination centre under the Pb-rich condition, there are no such centres with a low-energy formation energy under the Pb-poor condition~\cite{Li17apl}. There exist recent theoretical works of defect physics and chemistry for the hybrid layered perovskite (CH$_3$NH$_3$)$_2$Pb(SCN)$_2$I$_2$~\cite{Ganose17jmca}, bismuth-based lead free double perovskites~\cite{ZXiao16csc}, and defect passivation in the hybrid perovskites using quaternary ammonium halide anions and cations~\cite{Zheng17ne}.

\subsection{Ion diffusion}
Once the intrinsic ionic defects formed inside solids, they can migrate according to the minimum energy pathways, as the hybrid perovskites exhibit ionic charge transport as well as electronic conduction in experiments. Such ionic diffusion in the halide perovskites is closely related to the long-term stability and power conversion efficiency of PSCs. In particular, the anomalous hysteresis of the current-voltage (J-V) curves, which is a severe disadvantage of PSCs, and giant switchable photovoltaic effects, are importantly invoked by ion migrations~\cite{Richardson16ees}. The chemical decomposition of MAPI upon moisture exposure can also be explained partly by ion diffusion. In the simulations, the main tasks are to identify the migrating species and their migration pathways, and to quantitatively determine the underlying energetics, on condition that the formation of such ionic defects have been already clarified.

\begin{figure}[!th]
\begin{center}
~\includegraphics[clip=true,scale=0.4]{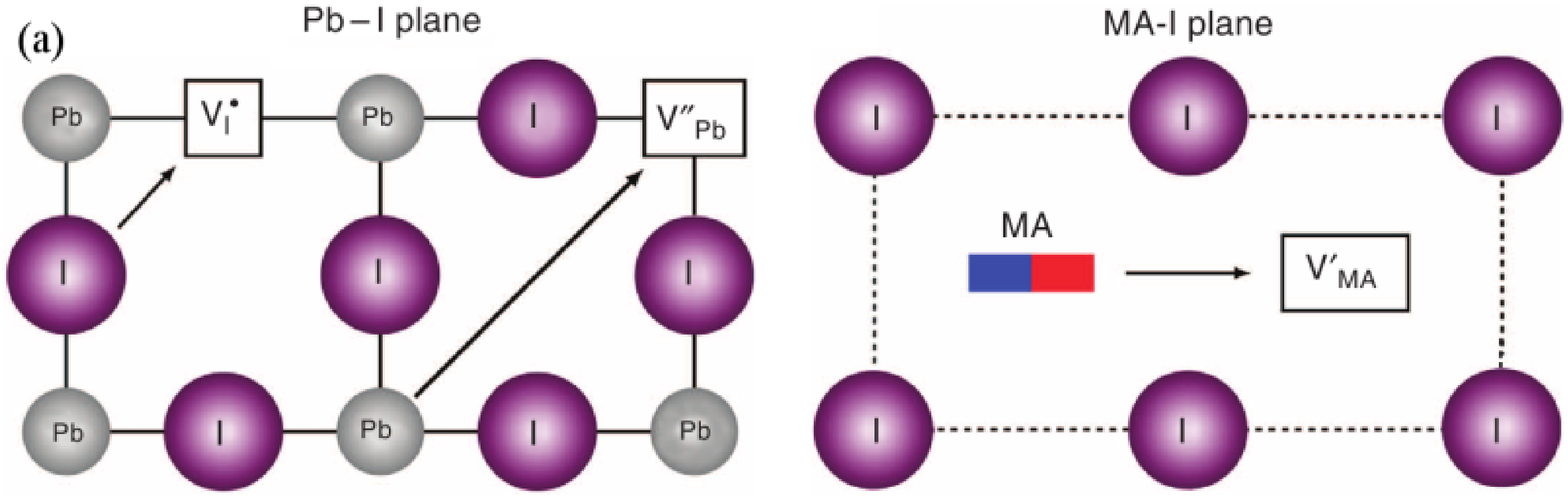} \\ \vspace{10pt}
\includegraphics[clip=true,scale=0.24]{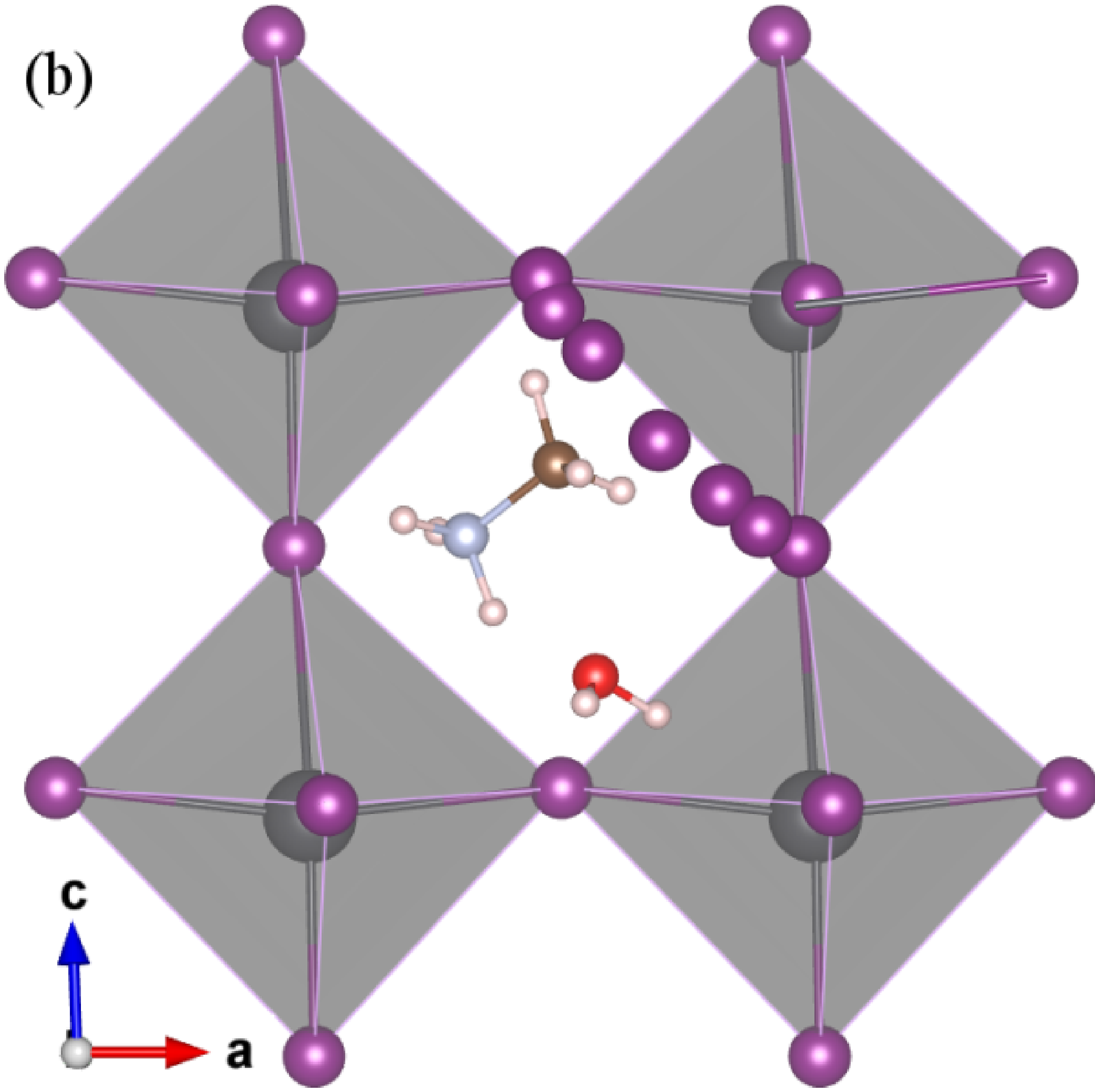}
\includegraphics[clip=true,scale=0.5]{fig13c.eps} 
\caption{\label{fig_ionmig}(a) Plausible pathways for vacancy mediated ion migrations of I$^{-}$, Pb$_{2+}$ and MA$^+$ in the hybrid iodide perovskite MAPI with a cubic phase. Adapted from~\cite{Eames15nc}. CC BY 4.0. (b) NEB images during I$^{-}$ migrations in the water-intercalated MAPI, and (c) activation energy profile for I$^{-}$ ion diffusion in the cubic MAPI, water-intercalated and mono-hydrated MAPI. Reproduced from~\cite{yucj18jmca}. CC BY 3.0.}
\end{center}
\end{figure}

Eames et al.~\cite{Eames15nc} investigated the diffusions of vacancy points (ions), not interstitials, in cubic MAPI, based on the fact that in ABX$_3$ halide perovskites vacancy-mediated ion diffusion is the most acceptable diffusion process while interstitial migration has not been observed in experiments. It should be noted that experiments probing ion migration, however, typically cannot distinguish between different migration pathways. They suggested three vacancy-mediated ion migration pathways including I$^-$ migration along the octahedron edge, Pb$^{2+}$ migration along the diagonal direction and MA$^+$ hopping into a neighbouring vacant site (Fig.~\ref{fig_ionmig}(a)). The activation energies for these ionic migrations were calculated to be 0.58, 2.31, and 0.84 eV, respectively~\cite{Eames15nc}. When compared with the measured values for hysteresis, the calculated value for I$^-$ is in good agreement with the measured value, 0.60$-$0.68 eV, whereas for Pb$^{2+}$ and MA$^+$ vacancies they are higher than the measured values, confirming that these ion migrations are difficult to occur~\cite{Eames15nc}. It should be noted that there is a wide spread of activation energies for ion migrations in both theory and experiments~\cite{Li17jpcm}.
Azpiroz et al.~\cite{Azpiroz15ees} considered three kinds of vacancies (V$_{\mtxt{I}}$, V$_{\mtxt{Pb}}$, V$_{\mtxt{MA}}$) and iodine interstitial (I$_i$) in tetragonal MAPI, emphasising that the migration of MA is responsible for the observed current-voltage hysteresis. The activation energies of these defects migrations were calculated to be 0.08, 0.46 and 0.80 eV along the similar pathways for three vacancies V$_{\mtxt{I}}$, V$_{\mtxt{MA}}$, V$_{\mtxt{Pb}}$ and 0.08 eV along the $c$ axis for I$_i$. In the case of MAPbBr$_3$, they are 0.09 and 0.56 eV for V$_{\mtxt{Br}}$ and V$_{\mtxt{MA}}$, respectively. Haruyama et al.~\cite{Haruyama15jacs} calculated the activation energies as 0.32/0.33 eV for V$_{\mtxt{I}}$/V$^{+}_{\mtxt{I}}$ and 0.57/0.55 eV for V$_{\mtxt{MA}}$/V$^{-}_{\mtxt{MA}}$ in tetragonal MAPI, 0.55/0.50 eV for V$_{\mtxt{I}}$/V$^{+}_{\mtxt{I}}$ and 0.61/0.57 eV for V$_{\mtxt{FA}}$/V$^{-}_{\mtxt{FA}}$ in trigonal FAPI, and 0.32/0.32 eV for Cl impurity I$_{\mtxt{Cl}}$/I$^{+}_{\mtxt{Cl}}$ in cubic MAPbI$_{3-x}$Cl$_x$. In these calculations, the common conclusion is that halide anions and MA or FA cations are the major ionic carriers due to their relatively low activation energies, which is consistent with the experimental findings and can explain the J-V hysteresis~\cite{Yuan16aem,Yuan15aem}. Based on the dilute diffusion theory, it was suggested that replacement of MA with larger cation FA can suppress the hysteresis and prevent the aging of PSC performance~\cite{Haruyama15jacs}.

Recently, we investigated the vacancy-mediated ion migrations in cubic MAPbX$_3$ (X = I, Br, Cl), and further water molecular diffusion in their water-intercalated and mono-hydrated phases, aiming at uncovering the role of water molecule in the chemical decomposition and thus the degradation of PSC performance~\cite{yucj18jmca}. It was found that the insertion of water into the MAPI reduces the activation energies of ion migrations (Fig.~\ref{fig_ionmig}(c)), indicating the easy decomposition of the hybrid halide perovskite under humid condition. Also, the activation barriers for ion and water migrations become higher from X = I to Br to Cl. These results indicate that the decomposition of halide perovskite occur through a multi-step process such as from the water intercalation to hydration and to decomposition, identifying the crucial role of water molecule in this process. Egger et al.~\cite{Egger15acie} investigated the hydrogen migration in the tetragonal MAPI to explain the hysteresis and material stability. Using the supercells containing a hydrogen impurity with different charge states, they found that the crystal structure may relaxed significantly for charged hydrogen impurities, collective iodide displacements can enhance proton diffusion, and the migration barriers for proton transfer are relatively low. Based on these findings, it was suggested that in the hybrid halide perovskites the hydrogen-like defects, introduced either extrinsically or intrinsically, may be mobile and play an important role in explaining the hysteresis effects and stability issues~\cite{Egger15acie}.

\section{Surface and interface}
The device performance and long-term stability of PSCs can be greatly affected by material processes occurring at surfaces, grain boundaries and interfaces as well. In reality, all of the bulk materials have the surface due to their finite size, and as such, the halide perovskite bulk crystals have several types of surfaces classified $hkl$-indices and terminations. Moreover, the perovskite films are fabricated mostly through solution processing methods, which make them polycrystalline and thus the formation of grain boundaries unavoidable. With respect to the device structure, it is necessary to adopt electron and/or hole transporting layers that contact to the photoabsorbing layer, halide perovskite, creating the interfaces such as TiO$_2$/MAPI for interface with electron transporting layer. Possibly important material processes at surfaces and interfaces are halide diffusion, ion accumulation, charge carrier transport and recombination at the defect states. For these phenomena, first-principles modelling and simulations can also provide useful microscopic insights as well.

\subsection{Surface phase diagram and electronic states}
In the standard simulation techniques using the three dimensional periodic boundary condition, the surfaces can be modelled by slab with a supercell, which consists of atomic layers and vacuum layer. The number of atomic layers and thickness of vacuum layer must be checked to ensure that the DFT total energy difference, e.g. surface formation energy, is not influenced by these modelling parameters. By applying {\it ab initio} atomistic thermodynamics, the formation energies of surfaces with various indices and terminations are calculated as functions of chemical potentials of constituent elements at finite temperature and pressure, producing a surface phase diagram.

The structural and electronic properties of the tetragonal MAPI surfaces with various indices and terminations have been investigated by using DFT calculations with the rev-vdW-DF functional and the inclusion of SOC effect~\cite{Haruyama14jpcl,Haruyama16acs}. Among low-index surfaces considered in that work, tetragonal (110) and (001) surfaces are flat nonpolar surfaces consisted of alternate stacking of the neutral [MAI]$^0$ and [PbI$_2$]$^0$ planes, while (100) and (101) surfaces are composed of charged [MAPbI]$^{2+}$ and [I$_2$]$^{2-}$ planes and [MAI$_3$]$^{2-}$ and [Pb]$^{2+}$ planes, implying large reconstruction or defect formation. Several types of PbI$_x$ polyhedron terminations were considered but not MA terminations. Identifying the constraint relations between the chemical potentials, the surface phase diagrams were drawn for the four types of these surfaces by calculating the Gibbs free energy differences as functions of chemical potentials of iodine and lead. Through these diagrams, stable terminations in each index surface can be determined at different growth conditions, i.e. Pb-rich (poor) and I-rich (poor) conditions. Also, the relaxed structures and electronic properties were analyzed in detail. Based on those calculations, it was concluded that a vacant termination is more stable than the PbI$_2$-rich flat termination on all the surfaces under thermodynamic equilibrium growth conditions of bulk MAPI. The flat terminations on the (001) and (110) surfaces were found to have surface states above the valence band of bulk (possibly attracting the photogenerated holes), which can be efficient intermediates of hole transfer to the adjacent hole transport materials with smaller energy loss~\cite{Haruyama14jpcl}.

\begin{figure}[!th]
\begin{center}
\includegraphics[clip=true,scale=0.45]{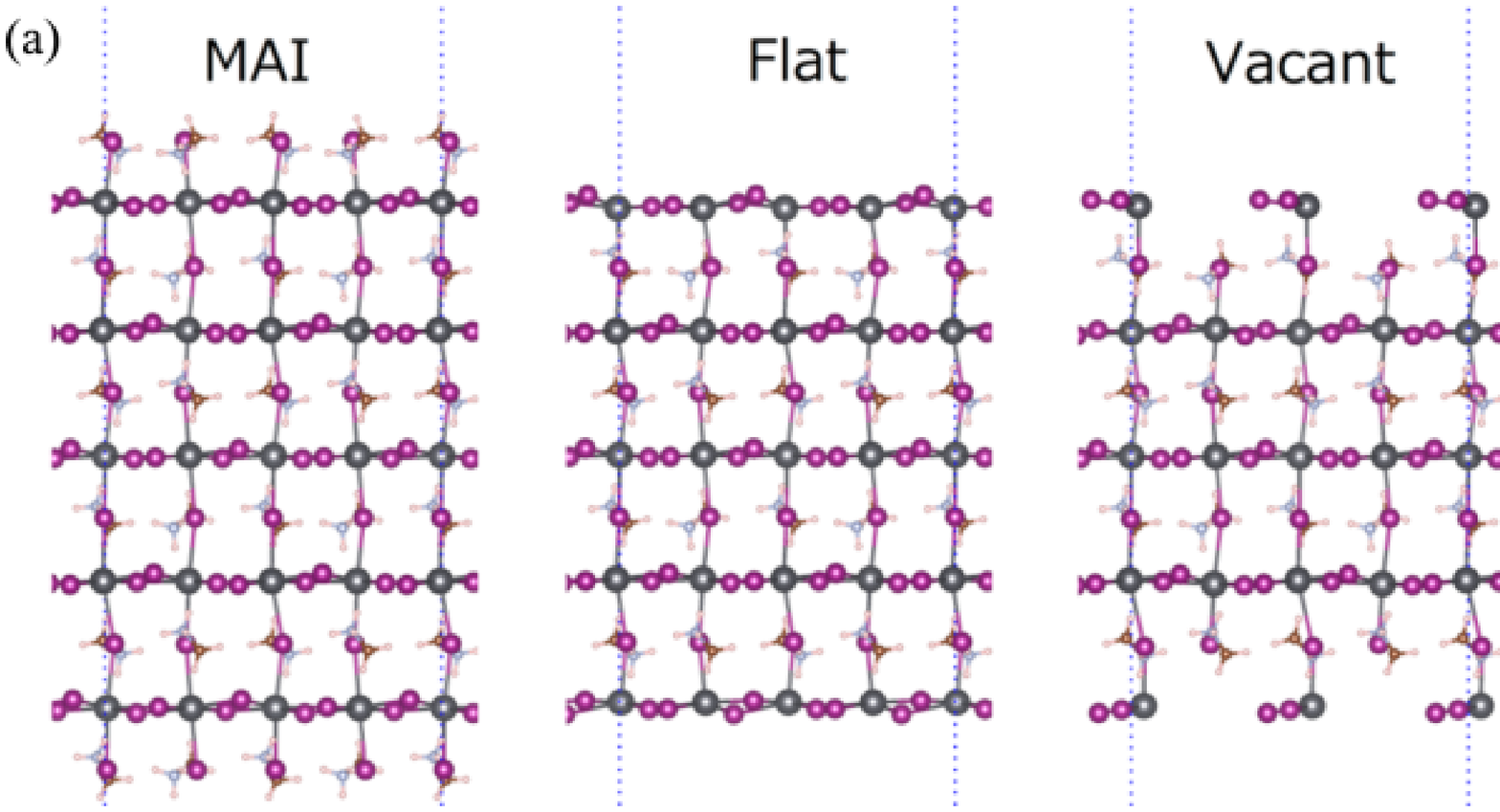} \\ \vspace{5pt}
\includegraphics[clip=true,scale=0.48]{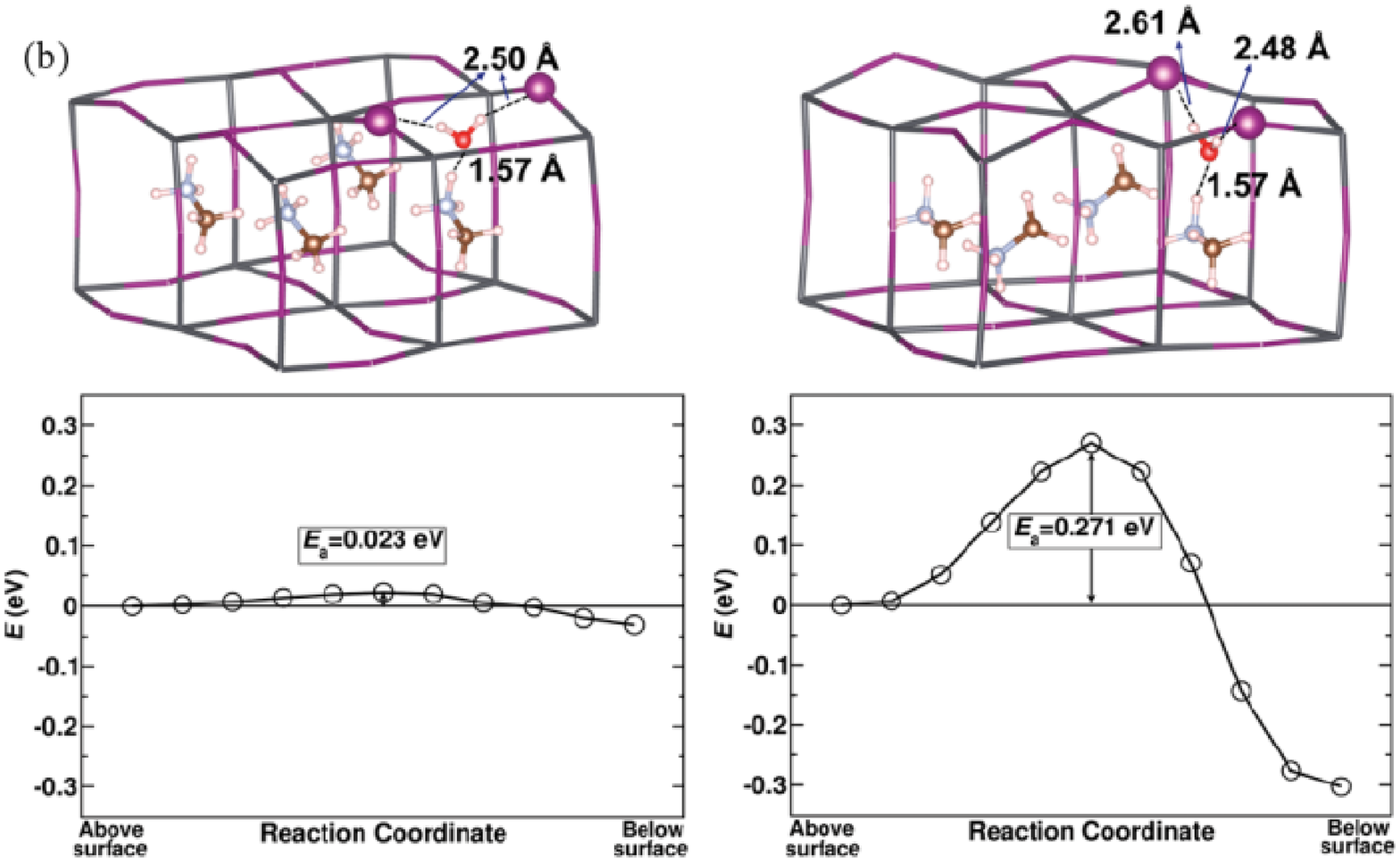}
\caption{\label{fig_surf}(a) Supercell models of tetragonal MAPI (001) surfaces with MAI, flat PbI$_2$, and vacant PbI$_2$ terminations, displaying cells by dotted lines. Reprinted with permission from~\cite{Uratani17jpcl}. Copyright 2017 American Chemical Society. (b) Lowest-energy structures of water under the first layer of the PbI$_2$-terminated $P^+$ surface (left) and of water initially placed at the hollow site of the PbI$_2$-terminated $P^-$ surface, and the corresponding reaction energy profiles. Reprinted with permission from~\cite{Koocher15jpcl}. Copyright 2015 American Chemical Society.}
\end{center}
\end{figure}
As discussed above, the trapping of charge carriers at defects on the surfaces and grain boundaries is one of the detrimental factors in PSC performance, because it causes nonradiative recombination loss, deteriorates the carrier lifetime, and originates the J-V hysteresis. Uratani and Yamashita~\cite{Uratani17jpcl} investigated the types of surface defects responsible for carrier trapping in tetragonal MAPI (001) surface by performing the HSE+SOC calculations. They constructed the ($2\times 2$) surface models with a vacuum region of 15 \AA~and three types of terminations, i.e. MAI, flat and vacant (Fig.~\ref{fig_surf}(a)). Several kinds of intrinsic point defects such as vacancies (V$_{\mtxt{I}}$, V$_{\mtxt{MA}}$, V$_{\mtxt{Pb}}$), interstitials (I$_i$, Pb$_i$) and antisites (Pb$_{\mtxt{I}}$, Pb$_{\mtxt{MA}}$) were identified to be formed on each terminated-surface, and their formation energies were calculated at I-rich, moderate and Pb-rich conditions. The energy levels of defect states were also drawn comparing with the VBM and CBM of each system. Using the calculation results, they concluded that under the I-rich condition iodine interstitials I$_i$ on flat and vacant surfaces are responsible for the carrier trapping, while under the Pb-rich condition V$_{\mtxt{I}}$ on vacant surfaces and Pb$_i$ defects are carrier traps. Therefore, the Pb-rich condition is better than the I-rich condition in terms of PV performance, which is consistent with the experiments.

Surface modelling and simulations are also important for understanding the interaction of halide perovskites with water, which can be related with the chemical decomposition of especially MAPI under humid condition. Such water-assisted chemical decomposition of MAPI can be explained by hydrolysis~\cite{Frost14nl} or hydration~\cite{yucj18jmca}. In these processes, the initial step is the adsorption of water on the MAPI surface and its diffusion (or penetration) into the bulk. Liu and co-workers explored the water effects on the detailed structure and properties of MAPI surfaces based on {\it ab initio} molecular dynamics using the Gaussian-type double-$\zeta$ polarized basis sets (DZVP) and the PBE+vdW XC functional~\cite{Tong15jpcl}. Tetragonal MAPI (001) surfaces with a MA$^+$ termination were considered as the base surfaces. It was found that the water adsorption energy on the MAPI (001) surface is about 0.30 eV and the diffusion barrier of water molecule from the surface to inside region is only about 0.04 eV, indicating that the water can easily penetrate into the bulk. Koocher et al.~\cite{Koocher15jpcl} found that the water adsorption is greatly affected by the orientation of the MA$^+$ cations close to the surface. The activation barriers of water penetrations were calculated to be 0.023 eV in the case of $P^+$ (CH$_3$-end of the MA molecule towards top surface) and 0.271 eV in the case of $P^-$ (NH$_3$-end) on the MAI-terminated (001) surface (Fig.~\ref{fig_surf}(b)). Therefore, it could be suggested that controlling orientation via poling or interfacial engineering can enhance the moisture stability. Choosing the MAI-terminated (110) surface in MAPI and the MABr-terminated (110) surface in MAPbBr$_3$, Zhang et al.~\cite{Zhang15jpcc,Zhang16jpcc} calculated the adsorption energies of water molecules to be 1.74 and 1.51 eV with PBE while 1.87 and 1.67 eV with PBE+vdW on MAPI and MAPbBr$_3$ surfaces. They suggested that the deprotonation of the MA cations followed by the desorption of the CH$_3$NH$_2$ molecules is a primary step in the degradation mechanism.

\subsection{Interface}
In PSCs, the halide perovskite photoabsorbers are sandwiched between the electron transport layer (ETL) and hole transport layer (HTL), so that the photoexcited electrons and holes can be efficiently transported to the collecting electrodes through these layers. In the case of MAPI, for example, it is in contact with mesoporous TiO$_2$ scaffold as ETL in one side and with organic Spiro-OMeTAD material as HTL in other side, thus creating heterojunction interfaces at these contacts~\cite{Butler16jmcc}. Such interfaces can be considered as the large defects, where nonradiative recombination of electrons and holes may occur, resulting in the decrease of PV performance. In fact, charge carrier extaction from the absorber to the transporter can be obstructed at the interfaces by several factors such as interfacial energy barriers originated from imperfect band alignment and charge carrier recombination driven by interfacial defect traps~\cite{Schulz18ael}. On the other hand, the interfaces can be weak against the infiltration of external particles such as water molecule and hydrogen atom or proton, which can trigger or facilitate the degradation of PSCs. Therefore, understanding the interface phenomena and further engineering the interface structure and composition are very important for improving the PSC performance.

Mosconi and co-workers investigated the interfaces of TiO$_2$ to MAPI and also MAPbI$_{3-x}$Cl$_x$ with a small Cl content ($\sim$4\%) using SOC-DFT calculations~\cite{Azpiroz15ees,Mosconi14jpcl,Roiati14nl}. The most stable tetragonal (110) and pseudocubic (001) surfaces of MAPI and MAPbI$_{3-x}$Cl$_x$ were selected and the model systems were made by $(3\times5\times3)$ supercells with slab stoichiometries MA$_{60}$Pb$_{45}$I$_{150}$ and MA$_{60}$Pb$_{45}$I$_{135}$Cl$_{15}$, which were fully optimized by performing atomic relaxations. These optimized supercells were deposited onto a $(5\times3\times2)$ slab of anatase TiO$_2$ (101) surface made by 120 TiO$_2$ units. The lattice mismatches were found to be +0.36 and +1.92\% for the (110) surfaces, and +0.75 and $-$1.85\% for the (001) surfaces, along the TiO$_2$ $a$ and $b$ directions, respectively. The binding of perovskite to TiO$_2$ substrate was found to be through the chemical bond between the perovskite halide atoms and the under-coordinated Ti atoms of the TiO$_2$ surface. From the simulation results, they conclude that MAPI and MAPbI$_{3-x}$Cl$_x$ tend to grow (110) surfaces on TiO$_2$ due to their higher binding energies to TiO$_2$ than (001) surfaces, and interfacial Cl atoms increase the binding energy of the MAPbI$_{3-x}$Cl$_x$ (110) surface compared to MAPI. Through the electronic structure calculations, it was revealed that the interaction of the perovskite with TiO$_2$ changes the interface electronic structure to a stronger interfacial coupling between the Ti $3d$ and Pb $6p$ conduction band states and to a slight upshift of TiO$_2$ conduction band energy~\cite{Mosconi14jpcl}. In order to investigate the effect of interface defects, vacancies V$_{\mtxt{I}}$ and V$_{\mtxt{MA}}$ were created at the perovskite surface exposed to the vacuum (HTL contact) and at the oxide-contacting surface, based on the already established fact that under working conditions V$_{\mtxt{I}}$ should diffuse towards the HTL while V$_{\mtxt{MA}}$ towards the ETL side~\cite{Azpiroz15ees}. When compared with the non-defective interface, the outermost valence band states of the perovskite intrude into the band gap of the TiO$_2$ substrate, while the conduction band states are found inside the manifold of the oxide's conduction states. The presence of defects at interfaces was found to modify the band alignment, cause the bending of the perovskite bands close to the TiO$_2$ surface, and create the trap states~\cite{Azpiroz15ees}. 

For the case of all-inorganic perovskites, CsPbBr$_3$/TiO$_2$ heterostructure has been investigated using the PBEsol+HSE06 functional~\cite{Qian18apl}. From the experimental observations for CsPbBr$_3$/TiO$_2$ interfaces, the interface modified CsBr and PbBr$_2$ layers were confirmed to be made during the synthesis, leading to the corresponding CsBr/TiO$_2$ and PbBr$_2$/TiO$_2$ interfaces. From the analysis of the charge population, it was found that large charge transfer occurs between Cs and O atoms, while much smaller charge transfer between Br and Ti atoms, being caused by the interface effects. Accordingly, the charge difference before and after the interface formation was drawn; the charge accumulation occurs on Ti and O atoms while the depletion on Cs and Br atoms, implying the electron transfer form perovskite to the TiO$_2$. From the integrated local density of states projected along the direction perpendicular to the interface, the staggered gap offset junction was observed for both interfaces, revealing the driving force of the charge transfer through the interfaces. Moreover, the band gap in the regions far away from the interface is almost consistent with that of individual CsPbBr$_3$ or TiO$_2$, whereas in the interface vicinity it decreases. The interface electronic states were found to locate on the conduction band edge of CsPbBr$_3$, indicating the shallow gap states. It was suggested that these shallow gap states may increase the absorbed photons and promote the optical absorption~\cite{Qian18apl}.

Regarding the degradation issue, the interaction between the MAPI surface and liquid water environment was investigated by making the slab interface model consisted of MAPI surface and water molecules and performing {\it ab initio} MD simulations~\cite{Mosconi15cm}. The perovskite surfaces were made by cutting $(2\times 2)$ slabs from the bulk tetragonal MAPI crystal, leading to (001) surfaces with three kinds of terminations: MAI, PbI$_2$ and PbI$_2$-defective terminations. To generate the defective surface, six PbI$_2$ units were removed out of a total eight PbI$_2$ units from the PbI$_2$-terminated (001) surface~\cite{Haruyama14jpcl}. The vacuum region was filled with water molecules, the numbers of which were determined from the experimental density of liquid water as 284, 226 and 235 for MAI-terminated, PbI$_2$-terminated and PbI$_2$-defective surfaces, respectively. By analyzing the MD trajectories, the MAI-terminated slabs were found to undergo rapid solvation according to a nucleophilic substitution of I by H$_2$O, desorption of MA molecules, and thus a net dissolution of a MAI molecule. On the contrary, the PbI$_2$-terminated surfaces were stable against such solvation process, although the percolation of a water molecule can occur to form a hydrated bulk phase. It was also concluded that PbI$_2$ defects may facilitate the solvation of the surface and initiate the degradation of the entire perovskite.

\section{Summary and outlook}
Being inspired by astonishing experimental findings for PSCs, a lot of computational works for the halide perovskites have been carried out within the latest five years in order to reveal the underlying mechanisms of perovskites' physics and chemistry and indicate a way for finding new perovskites with an improved performance. In this review, we have presented a critical discussion about the theoretical findings and conclusions for crystalline structures, electronic and optical properties, lattice dynamics and material stabilities, defect physics and ionic diffusions, and surfaces and interfaces in the hybrid organic-inorganic and purely all-inorganic halide perovskites, which have been made based on first-principles calculations. Specific theories, methods, and modelling approaches for these calculations have been provided, trying to point out the discrepancy and limit of the modelling and calculations when compared with the available experimental data. Special attention has been paid to making it clear the structure-property relationship in these materials, which was in close association with PV performance (efficiency and stability). We have demonstrated the predicting power of first-principles materials design with some predictions for new halide perovskites unexplored by experiment or optimal mixing ratios in some solid solutions.

In order to commercialize PSCs with a realistic impact on global energy market share, the efficiency and stability of PSCs need to be further improved, as the efficiency is still under the maximum theoretical efficiency of 25$\sim$27\% estimated by first-principles modelling~\cite{Granas16sr}. In particular, inherent chemical instability of halide perovskites under working conditions such as humid, light and thermal environment should be of the primary urgency in commercialization. On one hand, the performance enhancement can be realized by identifying the most ideal device architecture, as some materials-architecture combinations have shown high efficiency and stability. On the other hand, due to a rich versatility of halide perovskites with respect to the chemical composition and structural phase, it is desirable to find the most optimal choice of composition and structure that can give the best performance of PSC. Relying on computational or coupled experimental-computational method will be the best way to achieve this aim with respect to the time and cost.

Band gap engineering is the first thing to do. Once the chemical composition and crystalline structure are identified, the band gap can be calculated in almost autonomous way with tolerable wasting time and demand. However, much more time-consuming methods beyond standard DFT such as hybrid functional and $GW$ should be needed to get a reliable band gap and especially band alignment. Then, lattice dynamics should be performed to estimate the material stability at finite temperature and pressure. Although quasi-harmonic approximation can be used to get phonon dispersion, anharmonicity that the perovskites exhibit inherently should be considered; this is not yet well-established. Anharmonicity in lattice dynamics could be related to the ferroelectric nature of perovskite, which is still in debate. To save the time and cost, inexpensive interatomic potential functions can be used in lattice dynamics but with a careful consideration of accuracy. Finally, defect and interface engineering are also of great importance in the meaning of direct relation to synthesis process and device architecture. Despite the progress in these respects, many important questions remain untouched, e.g. proton percolation into the interface and influence on performance.

\section*{Author Information}
{\bf ORCID} \\
Chol-Jun Yu: 0000-0001-9523-4325 \\ \\
{\bf Notes} \\
The author declares no competing financial interest. \\ \\
{\bf Biography} \\
{\bf Chol-Jun Yu} holds a position as Head of Computational Materials Design group at Kim Il Sung University. With his group members, he focuses on design of energy materials such as perovskite solar cell materials, sodium ion battery materials, and ceramic composite coating materials for energy saving in buildings by using first-principles methods. He received his MSc. (2003) in Theoretical Physics from his home university and his Dr.rer.nat. (2009) in Computational Materials Engineering from RWTH Aachen University as a scholarship holder of Gottlieb Daimler and Karl Benz Foundation (2006--2009). He has a membership of International Association of Advanced Materials (IAAM) from 2017.

\section*{Acknowledgment}
I am profoundly grateful to Prof. Aron Walsh and Prof. C. Richard A. Catlow for kind invitation to TYC 5th Energy Workshop held in London, useful discussions during stay there, and positive recommendation for writing this Topical Review. This work was partly supported by the State Committee of Science and Technology, DPR Korea, under the fundamental research project ``Design of Innovative Functional Materials for Energy and Environmental Application'' (No. 2016-20).

\bibliographystyle{elsarticle-num-names}
\bibliography{Reference}

\end{document}